\documentclass[superscriptaddress,nofootinbib]{revtex4}
\usepackage{hyperref, doi, url}
\usepackage{bbm, amsfonts, amsmath,amssymb,mathrsfs}
\usepackage{graphicx}               
\graphicspath{{figs/}}
\usepackage[usenames,dvipsnames]{color}
\usepackage{amssymb}
\usepackage{epsfig,amsmath,graphics}
\usepackage{epstopdf}
\usepackage{subfigure}
\usepackage{verbatim}
\usepackage[normalem]{ulem}

\advance \topmargin by -\headheight
\advance \topmargin by -\headsep
\evensidemargin \oddsidemargin
\def\simgt{\mathrel{\lower2.5pt\vbox{\lineskip=0pt\baselineskip=0pt
           \hbox{$>$}\hbox{$\sim$}}}}
\def\simlt{\mathrel{\lower2.5pt\vbox{\lineskip=0pt\baselineskip=0pt
           \hbox{$<$}\hbox{$\sim$}}}}

\newcommand{\GeV}{\text{GeV}}
\newcommand{\eV}{\text{eV}}
\newcommand{\cm}{\text{cm}}
\newcommand{\Hz}{\text{Hz}}

\newcommand{\MHz}{\text{MHz}}
\newcommand{\GHz}{\text{GHz}}
\newcommand{\THz}{\text{THz}}
\newcommand{\mhidden}{m_{\gamma'}}
\begin{document}

\title{
A Radio for Hidden-Photon Dark Matter Detection
}

\author{Saptarshi Chaudhuri}
\affiliation{Department of Physics, Stanford University, Stanford, CA 94305}

\author{Peter W. Graham}
\affiliation{Stanford Institute for Theoretical Physics, Department of Physics, Stanford University, Stanford, CA 94305}

\author{Kent Irwin}
\affiliation{Department of Physics, Stanford University, Stanford, CA 94305}
\affiliation{SLAC National Accelerator Laboratory, Menlo Park, CA 94025}

\author{Jeremy Mardon}
\affiliation{Stanford Institute for Theoretical Physics, Department of Physics, Stanford University, Stanford, CA 94305}

\author{Surjeet Rajendran}
\affiliation{Berkeley Center for Theoretical Physics, Department of Physics, University of California, Berkeley, CA 94720}
\affiliation{Stanford Institute for Theoretical Physics, Department of Physics, Stanford University, Stanford, CA 94305}

\author{Yue Zhao}
\affiliation{Stanford Institute for Theoretical Physics, Department of Physics, Stanford University, Stanford, CA 94305}

\begin{abstract}
We propose a resonant electromagnetic detector to search for hidden-photon dark matter over an extensive range of masses.
Hidden-photon dark matter can be described as a weakly coupled ``hidden electric field,'' oscillating at a frequency fixed by the mass, and able to penetrate any shielding.
At low frequencies (compared to the inverse size of the shielding), we find that the observable effect of the hidden photon inside any shielding is a real, oscillating magnetic field.
We outline experimental setups designed to search for hidden-photon dark matter,
using a tunable, resonant LC circuit designed to couple to this magnetic field.
Our ``straw man'' setups take into consideration resonator design, readout architecture and noise estimates.
At high frequencies, there is an upper limit to the useful size of a single resonator set by $1/\nu$.
However, many resonators may be multiplexed within a hidden-photon coherence length to increase the sensitivity in this regime.
Hidden-photon dark matter has an enormous range of possible frequencies, but current experiments search only over a few narrow pieces of that range.
We find the potential sensitivity of our proposal is many orders of magnitude beyond current limits over an extensive range of frequencies, from 100 Hz up to 700 GHz and potentially higher.
\end{abstract}

\maketitle
\vspace{-10pt}

\tableofcontents

\section{Introduction}

The astronomical, astrophysical and cosmological evidence for dark matter provide a tantalizing hint of physics beyond the Standard Model. This evidence has motivated decades of concerted effort to detect or create dark matter. This search has principally focused on Weakly Interacting Massive Particles (WIMPs), and supersymmetry, which are very well motivated. However, there is no guarantee that either WIMPs or supersymmetry exist.  Furthermore, the limits from the LHC on new weak-scale physics and the lack of discovery of WIMPs to date after decades of direct detection experiments strongly suggest that other candidates should be investigated.

Since the model of dark matter is not known, it is important to search for broad classes of dark matter candidates. WIMP direct detection searches are powerful probes not only of weakly-interacting candidates, but also a broad class of particles with masses several orders of magnitude from the weak-scale in either direction and any generic coupling to the fermions of the Standard Model.  Although WIMP detectors use many different technologies, they are all optimized for the detection of the energy deposited by the scattering of a single dark matter particle with either the nucleus or an electron in the detector.  This detection strategy is appropriate since WIMPs are relatively massive, with low phase-space density and particle-like behavior.

However, another generic class of dark matter candidates exhibits field-like behavior. A light (bosonic) field making up the dark matter with mass $\ll 0.1$ eV will have a high phase-space density, since the local dark matter density is $\rho_\text{DM} \sim 0.3 \, \GeV / \cm^3 \sim (0.04 \, \eV)^4$.  Such light-field dark matter is generally produced non-thermally (unlike the WIMP), often by the misalignment mechanism, and is best described as a classical, background field oscillating with frequency roughly equal to its mass \cite{Dine:1982ah, Preskill:1982cy, Abbott:1982af}.  There are only a few general types of dark matter candidates (consistent with effective field theory) in addition to WIMPs and light fields, including topological dark matter \cite{Pospelov:2012mt, Derevianko:2013oaa} and ultra-heavy candidates such as primordial black holes.  Light field dark matter represents a broad class of dark matter candidates, and an extremely attractive target for dark matter searches.

Searching for the energy deposition from a scattering event is not a promising technique to constrain such light fields.  A more promising strategy is to search for the coherent interactions of that field (akin to forward scattering), in order to overcome the likely very weak couplings and obtain an observable signal.  Axion detectors such as ADMX \cite{Duffy:2006aa, Asztalos:2011bm, Sikivie:1983ip} and the proposed CASPEr \cite{Budker:2013hfa} are examples of this type of strategy.  These experiments constrain not only models of the QCD axion, but also models with a generic scalar with any of the couplings allowed for a pseudo-Goldstone boson \cite{Graham:2011qk, Graham:2013gfa}.  These detectors cover the scalar case well, but the other well-motivated and relatively unexplored possibility is a vector field\footnote{A background spin 2 field is also possible, with the most prominent example being a gravitational wave (though not as dark matter).  These detectors necessarily search for coherent, classical field-like interactions of the gravitational wave instead of hard, particle-like scattering for the same reasons as the light-field DM detectors we are discussing \cite{Dimopoulos:2008sv, Graham:2012sy}.} \cite{Nelson:2011sf, Arias:2012az}.  Light-through-walls type experiments using microwave cavities \cite{Ahlers:2007rd, Jaeckel:2007ch, Povey:2010hs, Parker:2013fba, Betz:2013dza, Graham:2014sha} can search for a new vector that exists in the theory, but do not indicate whether it is dark matter.  There has been much recent interest in hidden-sector photons, and limits come from astrophysical production and collider experiments \cite{Pospelov:2007mp, Abel:2008ai, ArkaniHamed:2008qn, ArkaniHamed:2008qp, Pospelov:2008zw, Goodsell:2009xc, Arvanitaki:2009hb, Jaeckel:2010ni, Essig:2010ye, Ringwald:2012hr, Reece:2009un, Batell:2009di, Bjorken:2009mm, Aubert:2009af, deNiverville:2011it, Hewett:2012ns, Dharmapalan:2012xp, Battaglieri:2014hga}.  But there are only a few possible experiments to search for hidden-sector photon dark matter, including ADMX \cite{Wagner:2010mi} and the dish proposal \cite{Horns:2012jf, Dobrich:2014kda}, and they can only search in a certain range of masses.  It is thus important to find ways to cover other pieces of the mass spectrum, since there is no sharp prediction for the mass of hidden photon dark matter.

Such light fields can arise naturally from high energy physics.  The axion can arise from UV physics, for example it is generic in string theories.  Another natural possibility is to have additional hidden sectors, e.g.~new gauge groups.  String theory, for example, often produces extra $U(1)$'s.  While such a sector may be complicated, we can parametrize all effects by the effective operator coupling it to the Standard Model.  Besides a direct coupling (i.e. charging the Standard Model under this gauge symmetry), the only other generic possibility is a kinetic mixing between the new $U(1)$ and electromagnetism \cite{Holdom:1986eq}, the hidden-photon.  Because this is a dimension 4 operator it can arise at very high energies and will still have observable effects in low energy experiments.  Furthermore, this coupling covers a wide range of possible models of new physics, and is one of very few relevant, dimension 4 couplings that are allowed between the Standard Model and any new dark matter sector. Such an experiment is thus promising and well motivated.

\section{Overview}
\label{sec:overview}

Hidden photons are massive $U(1)$ vector bosons that kinetically mix with the standard model. After rotating to the ``mass basis'', they are described by the following Lagrangian\footnote{For a more careful treatment of hidden photons and their interaction with electromagnetic systems see~\cite{Graham:2014sha}, building on (and correcting) earlier treatments~\cite{Ahlers:2007rd,Jaeckel:2007ch}.}:
\begin{equation}
\mathcal L \supset -\frac{1}{4} \left( F_{\mu \nu} F^{\mu \nu} + F'_{\mu
\nu} F'^{\mu \nu} \right) + \frac{1}{2}\mhidden^2 A'_{\mu} A'^{\mu}
- e J_{EM}^{\mu} \left( \, A_{\mu} + \varepsilon \, A'_{\mu}\right)
\label{eq:mass-basis-Lagrangian}
\end{equation}
where $A_{\mu}$ and $F_{\mu \nu}$ represent the gauge potential and field strength of electromagnetism, $A'_{\mu}$,  $F'_{\mu \nu}$ and $\mhidden$ represent the gauge potential, field strength and mass of the  hidden photon, $J_{EM}^{\mu}$ is the electromagnetic current and $\varepsilon$ is the small kinetic mixing parameter. 
In vacuum, the hidden-photon field obeys the wave equation $(\partial_t^2 -\nabla^2+\mhidden^2)A'_\mu = 0$, with the constraint $\partial_\mu A'^\mu = 0$.

The hidden photon can be understood intuitively as a new particle that behaves like the regular photon, except that \emph{a}) it has a mass, and \emph{b}) it interacts only weakly with charged particles, with coupling suppressed by $\varepsilon$.
In simple terms this has three important consequences for us. 
Firstly, the mass allows hidden photons to behave as cold matter, and thus to be considered as a dark matter candidate~\cite{Nelson:2011sf}. 
Secondly, the small coupling to charged particles (specifically electrons) means that hidden photons can weakly excite electromagnetic systems. 
Thirdly, with small coupling and macroscopic Compton wavelength, hidden photons have an extremely long penetration depth in conductors (and superconductors), and so do not get screened by electromagnetic shielding\footnote{We note that, since dark-matter hidden photons are highly non-relativistic, the distinction between longitudinal and transverse modes is unimportant and does not affect their ability to penetrate shielding, unlike with relativistic hidden photons~\cite{An:2013yfc, Graham:2014sha}.}.

It can be helpful to rewrite the lagrangian in the ``interaction basis'', related by $A_\mu + \varepsilon A'_\mu = \tilde A_{\mu}$ and $A'_\mu - \varepsilon A_\mu = \tilde A'_\mu$, giving
\begin{equation}
\mathcal L \supset -\frac{1}{4} \left( \tilde F_{\mu \nu} \tilde F^{\mu \nu} + \tilde F'_{\mu \nu} \tilde F'^{\mu \nu} \right)
 + \frac{1}{2}\mhidden^2 \tilde A'_{\mu} \tilde A'^{\mu}
- e J_{EM}^{\mu}  \tilde A_{\mu} + \varepsilon \mhidden^2 \tilde A_{\mu} \tilde A'^{\mu} \, .
\label{eq:int-basis-Lagrangian}
\end{equation}
This puts the coupling between the hidden photon and the Standard Model in the form of a small mass mixing $\varepsilon m_{\gamma'}^2$ with the photon. 
The smallness of this mixing results in an enormous penetration depth for the hidden photon in any material. 
It can be seen easily from Eq.~\ref{eq:int-basis-Lagrangian} that, in the limit of no back-reaction, a background hidden photon field $\tilde A'^\mu$ is equivalent to a effective current density $J_{EM, \rm eff}^\mu = - \varepsilon m_{\gamma'}^2 \tilde A'^\mu$. 

To gain intuition about hidden-photon dark matter, consider how the photons in a laser contribute coherently to make a classical electric field of the form $\vec E \approx \vec E_0 \cos (\omega t - \omega z)$.
The large number density of cold hidden photons comprising the dark matter contribute coherently to a classical ``hidden-electric'' field of the form $\vec E' \approx \vec E_0' \cos(\mhidden t)$. $E'_0$ is set by the local dark matter density to be $\sim\! 100 \, \eV / \cm$.
It pervades space since it is not affected by electromagnetic shielding, while its frequency is fixed to (very close to) $\mhidden$ because it is highly non-relativistic.

Given this intuition, one general scheme for detecting hidden-photon dark matter is to place a resonant electromagnetic detector inside an electromagnetic shield. The hidden-photon field will penetrate the shield and can weakly excite the detector.
An important subtlety in such setups is the effect of the shield itself on the detected signal.
While the hidden photon will not be blocked by the shield, it will nevertheless move charges in it, and their motion will in turn create new electromagnetic fields inside the shield. These fields have the potential to cancel the signal from the hidden photon itself, and so it is essential to treat the effect of the shield carefully.

\begin{figure}[t]
\begin{center}
\includegraphics[width=1.0\textwidth]{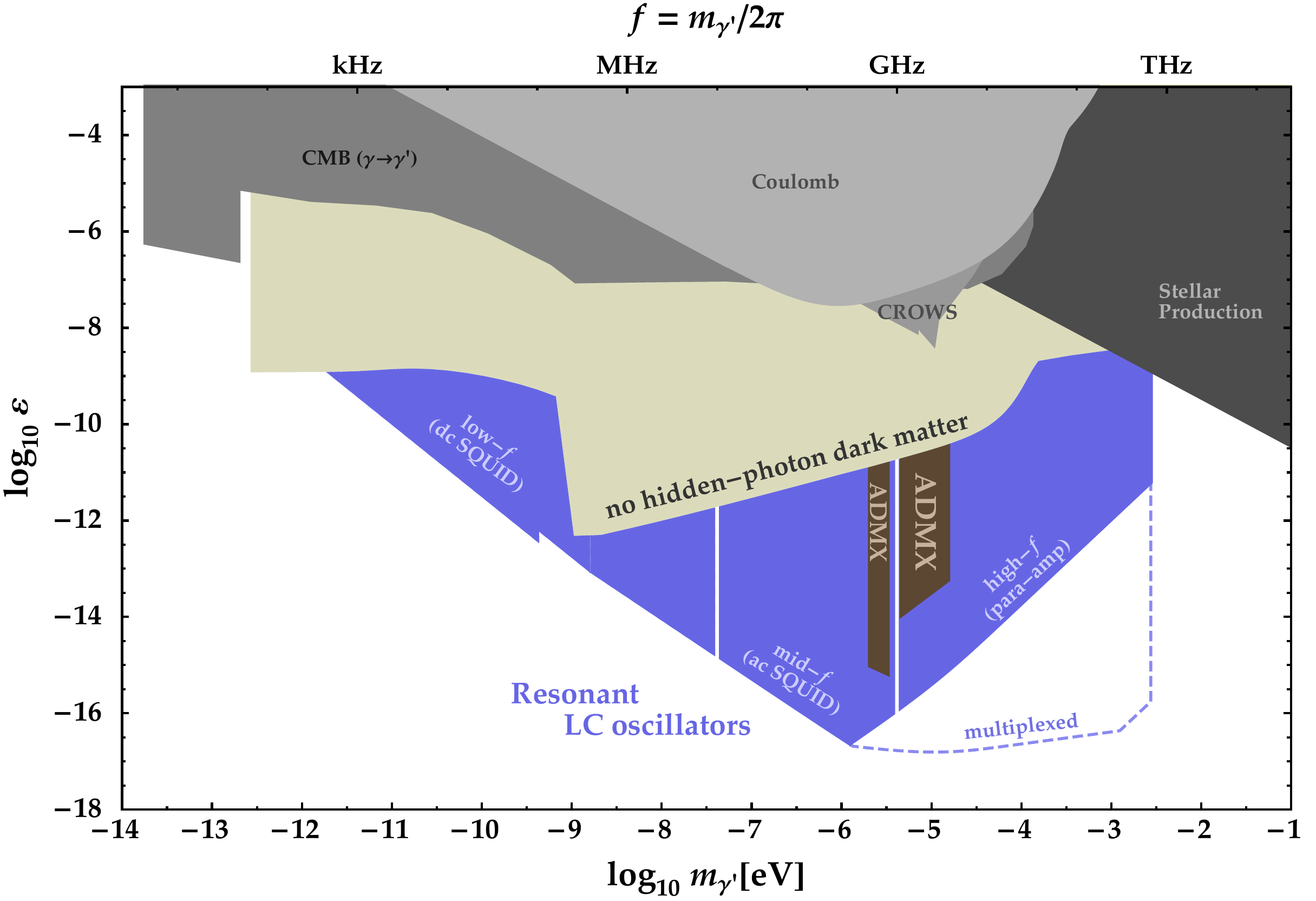}
\caption{Estimated sensitivity for hidden-photon dark matter searches with LC-oscillators.
The blue regions show the expected reach with the setups described in section~\ref{sec:strawman}.
The pale blue dashed line shows the improvement in reach that can be potentially achieved by multiplexing many detectors within one coherence length. See section~\ref{sec:sensitivity} for more details.
Hidden photons cannot make up all of the dark matter in the tan-shaded region~\cite{Arias:2012az}, and are excluded entirely by various constraints in the gray shaded regions~\cite{Jaeckel:2010ni, An:2013yfc, Graham:2014sha}. The narrow brown bands around GHz frequencies show the approximate region excluded by the ADMX experiment~\cite{Arias:2012az}.
} \label{fig:sensitivity}
\end{center}
\end{figure}

In this paper, we propose a setup in which the detector is a tunable high-$Q$ LC circuit coupled to a SQUID magnetometer or parametric amplifier. This setup will allow sensitivity to hidden-photon dark matter over an enormous range of masses.
The estimated reach of this setup is shown in figure~\ref{fig:sensitivity}.
We motivate and outline this setup in section~\ref{sec:resonantLC}, and present the signal size expected after accounting for the effect of the shielding (which is treated carefully in appendix~\ref{sec:signal-in-shield-full-calc}).
In section~\ref{sec:strawman} we outline preliminary designs for implementing this concept over a wide range of frequencies, from $\sim\! 100 \,\Hz$ to $\sim\! \THz$. This includes considerations of the resonator design, readout architecture, and dominant noise sources.
In section~\ref{sec:sensitivity} we present the estimated sensitivity of these setups to hidden-photon dark matter.
Finally, we conclude in section~\ref{sec:conclusions}.
First, however, we proceed in section~\ref{sec:HPDM} with a more detailed discussion of hidden-photon dark matter itself.

\section{Hidden photon dark matter}
\label{sec:HPDM}

A hidden photon with a small mass and a sufficiently small kinetic mixing $\varepsilon$ behaves as non-interacting, stable, cold matter.
It therefore makes a good phenomenological dark matter candidate (by which we mean that if the correct abundance of cold hidden photons can be produced prior to matter-radiation-equality, with the correct adiabatic density fluctuations, then it behaves just as it should from then on).
With a small mass ($10^{-22} \,\eV \ll \mhidden \ll \eV$), the high occupation number needed for hidden photons to make up all the dark matter density ensures that they behave as a classical field (in the same way as axion dark matter).
Hidden photon dark matter was first proposed in~\cite{Nelson:2011sf}, and investigated more thoroughly in~\cite{Arias:2012az}.
A more natural mechanism for generating a hidden-photon abundance in the early universe was presented in~\cite{Graham:2015rva}. 
For this paper, we simply assume that some mechanism has caused the hidden-photon field to have the correct ``initial condition'' at some point prior to matter-radiation-equality.

The hidden photon field obeys the wave equation
$(\partial_t^2 -\nabla^2+\mhidden^2)A'_\mu = 0$.
This means that for non-relativistic hidden-photon dark matter, the 3-vector field $\vec A'$ can be written in the form
\begin{align}
\vec A'(\vec x, t)  &=  A' e^{i \mhidden t} \int d^3 v \, \vec \psi (\vec v) e^{i \mhidden (\frac{1}{2} v^2 t - \vec v \cdot \vec x)}
\nonumber \\
&= A' e^{i \mhidden t} \times \hat n(\vec x, t) \, e^{i \varphi(\vec x, t)} \, .
\end{align}
$\hat n$ is a unit vector specifying the direction of the field.
The 0-component $A'_t$ is fixed by the constraint $\partial_t A'_t=\vec \nabla \cdot \vec A'$, and is small.
Note that this field is almost all hidden-electric, $\vec E' \approx - i \mhidden \vec A'$, while the hidden-magnetic field is velocity suppressed, $B' \sim v E'$.
The energy density in the field (in the non-relativistic limit) is given by
\begin{equation}
\rho = \frac{1}{2} \big( |\dot {\vec A}'|^2 + \mhidden^2 |\vec A'|^2 \big) = \frac{1}{2} \mhidden^2 A'^2 \, .
\end{equation}
Assuming the hidden photon makes up all of the dark matter, our local dark matter density of $\sim\! 0.3 \, \GeV / \cm^3$ therefore sets the amplitude of the field to be
\begin{equation}
|\vec E'| = \mhidden A' = \sqrt{2 \rho_{\rm DM}} \sim 100 ~\text V / \cm
\end{equation}
(while the magnetic component $B'$ is $\sim\! 10^{-8}\,$T).

The oscillation frequency is set by the hidden-photon mass,
\begin{equation}
\nu=m_{\gamma'}/2\pi \approx 2.5 \,{\rm MHz}\,\times (m_{\gamma'}/10^{-8} \,{\rm eV}) \, .
\end{equation}
The small but non-zero velocity spread of our local dark matter (encoded in $\vec \psi(\vec v)$) causes a small frequency spread, or equivalently a finite coherence length and time for the phase of the field $\varphi$. For standard DM velocity profiles, $v \sim \Delta v \sim 10^{-3} c$, resulting in\begin{gather}
\delta \nu / \nu \approx v^2 \approx 10^{-6}
\\
t_{\rm coherence} \approx  \frac{1}{\nu \, v^2} \approx 0.4 \, \text{s} \times (10^{-8} \eV/\mhidden)
\\
\lambda_{\rm coherence} \approx \frac{1}{\nu \, v} \approx 100 \, \text{km} \times (10^{-8} \eV/\mhidden) \, .
\label{eq:coherence-length}
\end{gather}

The direction of the field, $\hat n$, will also vary slowly over space and time. The coherence length and time of $\hat n$ must be \emph{at least} as long as that for the phase, and may be much longer (as discussed in~\cite{Arias:2012az}, this is a complicated question depending on the dynamics of structure formation).

The small frequency spread and sizable amplitude of the hidden photon field lend themselves to laboratory searches with resonant electromagnetic detectors. There are precision sensors suitable for operation over a wide frequency range, which may be deployed to detect the random classical fields of the hidden-photon dark matter.
The rest of the paper is devoted to a proposal for implementing this idea using resonant LC circuits.

One point worth making in passing is that the macroscopic spatial coherence of the field, in both its phase and direction, are predictions which may be eventually be tested, for example by running multiple experiments simultaneously, with different orientations and in different locations. This may be crucial in confirming that a tentative signal is truly hidden photon dark matter.

As a final comment, it should be noted that the magnitude of the $E'$ fields is independent of the hidden photon mass $\mhidden$. This is because the energy density in the field is assumed to be set by an initial condition instead of it being produced through standard model interactions (in which case, the production vanishes as $\mhidden \rightarrow 0$).  However, even though the fields can have a non-zero amplitude when $\mhidden \rightarrow 0$, they cannot be a dominant component of the dark matter if their mass is below $\sim\! 10^{-22}$~eV, since their de-Broglie wavelength would be larger than the size of the galaxy~\cite{Hu:2000ke}.

\section{A resonant search with an LC circuit}
\label{sec:resonantLC}

As we have seen in section~\ref{sec:overview}, hidden-photon dark matter is an oscillating field that pervades space and has a small coupling to electric charges and currents.
An immediate consequence is that any type of electromagnetic resonator will typically be excited by this ambient field, if its resonant frequency matches the hidden photon's oscillation frequency.
The size of the coupling between the field and the resonator is of course suppressed at least by a factor of $\varepsilon$.
On the other hand, a high-$Q$ resonator will ring up over many cycles, up to a maximum of around $10^6$, set by the coherence time of the hidden-photon field (see section~\ref{sec:HPDM}).
Searching for a small but unshieldable excitation, which appears in electromagnetic resonators only when tuned to one particular frequency, is therefore an ideal way to detect hidden-photon dark matter.

As was observed in Ref.~\cite{Arias:2012az}, a search of exactly this type is already being carried out by ADMX (the Axion Dark Matter eXperiment)~\cite{Asztalos:2009yp}. The electromagnetic resonator used by ADMX is a tunable microwave cavity, 
placed in a large static $B$-field to enable conversion of an axion dark matter field to an electric field. For hidden-photon dark matter, this $B$-field plays no role and does not affect the signal. Ref.~\cite{Arias:2012az} reinterpreted the axion search results from ADMX (and precursor experiments) to place limits on hidden-photon dark matter of $\varepsilon \simlt 10^{-14}$ in a mass window around $10^{-6}$$-$$10^{-5}$~eV, corresponding to a frequency range around 0.3$-$3~GHz.

While such cavity-based searches are highly sensitive and have the advantage of leveraging existing experiments, they are only useful in a limited mass range. In particular, since a cavity's lowest frequency scales inversely with its size, they are impractical for probing masses much below around $10^{-6}$~eV (corresponding to meter-scale cavities).

To search for hidden-photon dark matter in the enormous open parameter space below $10^{-6}$~eV, we propose using a tunable LC circuit in place of a microwave cavity.\footnote{For related proposals for axion dark matter searches see~\cite{Scott:talk, Sikivie:2013laa}.} LC circuits are resonators with a frequency $\omega = 1/\sqrt{L C}$, and hence low frequencies are reached by using large inductances or capacitances.
(This can be achieved by using many-turned inductors or small-separation capacitor plates, rather than by using geometrically large components.)

Unlike a microwave cavity, whose conducting walls naturally self-shield it from external fields, an LC circuit is highly sensitive to external non-dark-matter signals (it is essentially just a radio). To eliminate these, the LC circuit must therefore be placed inside a conducting shield, into which the only field that can possibly penetrate is the hidden-photon field. As we analyze in detail in appendix~\ref{sec:signal-in-shield-full-calc}, the conducting walls of the shield have a significant effect on fields found inside it. Conduction electrons in the shield experience a force from the $E'$-field associated with the hidden-photon dark matter. They respond by rearranging so as to source an $E$-field which cancels the net force on them. This cancellation occurs within the conducting walls, but is also effective inside the shield, so that the \emph{observable} field $\vec E_{obs}=\vec E + \varepsilon \vec E'$ is highly suppressed, with a parametric size
\begin{equation}
\qquad\qquad\qquad E_{obs} \sim \varepsilon \,  \sqrt{\rho_{\rm DM}} e^{i m_{\gamma'} t}
\times \big( (m_{\gamma'} R)^2 + (m_{\gamma'} R v_{\rm DM}) \big)
\qquad\qquad \text{(highly suppressed)} \, ,
\label{eq:E-obs}
\end{equation}
where $v_{\rm DM} \approx 10^{-3}$ is earth's speed with respect to the dark matter halo, $R$ is the characteristic size of the shield, and $m_{\gamma'}R$ is assumed to be small.
In fact, it is the $B$-field, generated by the motion of the conduction electrons in the shield walls, that dominates inside the shield. In the case that the shield is cylindrical and is aligned with the direction of $\vec A'$, $\vec B_{obs}$ is given by (see Eq.~\ref{eq:B-in-cylinder})
\begin{equation}
\qquad\qquad\qquad\qquad\qquad\qquad \vec B_{obs} \approx -\varepsilon \, \sqrt{\rho_{\rm DM}} e^{i m_{\gamma'} t} \hat \phi
\times m_{\gamma'} r
\qquad\qquad\qquad \text{(dominant observable field)} \, .
\label{eq:B-signal}
\end{equation}
Again, $m_{\gamma'}R$ is assumed to be small here.

Taking this result, the basic experiment design we propose is as follows. A high-$Q$ tunable LC circuit is placed inside a conducting shield, with the inductor wrapped around the $\phi$-direction, so as to couple to the driving $B$-field of Eq.~\ref{eq:B-signal}. The inductor is to be as large as possible to maximize the signal power and hence the signal-to-noise ratio. If the resonant frequency of the circuit is correctly tuned to the hidden-photon oscillation frequency, \begin{equation}
\nu=m_{\gamma'}/2\pi \approx 2.5 \,{\rm MHz}\,\times (m_{\gamma'}/10^{-8} \,{\rm eV}) \, ,
\end{equation}
the average field in the inductor will ring up after $Q\sim 10^6$ cycles to a size
\begin{align}
B_{\rm sig} &\approx Q \, \varepsilon \sqrt{\rho_{\rm DM}} \times \nu V_{\rm ind}^\frac{1}{3}\\
&\approx 3\times10^{-14} \, {\rm T} \times \left(\frac{Q}{10^6}\right) \left(\frac{\varepsilon}{10^{-12}}\right) \left(\frac{\nu}{\rm MHz}\right)  \left(\frac{V_{\rm ind}}{\rm m^3}\right)^\frac{1}{3} \sqrt{\frac{\rho_{\rm DM}}{0.3\,{\rm GeV}{\rm cm}^{-3}}} \, ,
\label{eq:B-signal}
\end{align}
where $V_{\rm ind}$ is the volume of the inductor.
This signal is to be read out with a SQUID or parametric amplifier coupled to the LC circuit.
Scanning through frequencies in increments of $\delta \nu \approx \nu/Q$ will allow many decades in hidden-photon mass to be explored.

In general, of course, the orientation of $\vec A'$ with respect to the experiment is unknown changes over time. This will have an $\mathcal O(1)$ effect on the size and direction of the $B$-field. For the purposes of this proposal we ignore these details, and simply note that $\vec A'$ does not change direction while the signal is ringing up, and neither does the earth rotate significantly. Using Eq.~\ref{eq:B-signal} will therefore give approximately the correct signal size in all but extremely rare cases of unlucky misalignment. (We refer interested readers to Eq.~\ref{eq:B-full-solution} for the result in full generality.)
In the future, testing this orientation-dependence may be a useful cross check if a signal is observed.

We will also consider setups to search at higher frequencies, for which $\nu^3 V_{ind}>1$.
At these frequencies, the $\mhidden R$ or $\nu V_{\rm ind}^{1/3}$ ``suppression'' factors in Eqs.~\ref{eq:E-obs} and~\ref{eq:B-signal} are no longer small, and the parametric size of the observable fields (before resonant enhancement) is
\begin{align}
\qquad\qquad\qquad\qquad
E_{\rm obs} \sim B_{\rm obs} \sim  \varepsilon \sqrt{\rho_{\rm DM}}
\qquad\qquad\qquad\qquad
\text{(high-frequency regime).}
\end{align}
We note that these fields will vary spatially on a scale $\sim\!\nu^{-1}$, which will also be the maximum useful size of a single LC oscillator designed to pick up the fields.
However, these fields will be coherent over the much larger distance given by Eq.~\ref{eq:coherence-length}. The signals of any individual LC oscillators placed within this distance may therefore be added coherently to achieve an effective volume much larger than $\nu^{-3}$. We defer a more detailed study of the interior fields in this regime, and the optimal resonator geometries to detect them, to future work.

\section{Description of experimental setup}
\label{sec:strawman}

In this section, we consider constraints on practical experimental designs, and establish a ``straw-man" implementation, including shielding, resonator design, and readout scheme. We evaluate the achievable sensitivity in the context of this straw man. We show that optimal resonator Q
can in principle be achieved in an experiment that is limited only by the thermal noise in frequency range 100 Hz-1 GHz and limited by quantum noise of both the cavity and the amplifier in the frequency range 1 GHz-700 GHz. However, practical limitations on parasitic resonances and inductor size may limit the performance at both the highest and the lowest frequencies.

\subsection{Constraints on Experimental Design}

We begin by describing design constraints on both the LC resonator and the readout circuit. As illustrated in the previous section, the magnetic field induced by the photon field in the shield is circumferential. We construct an LC resonant circuit that efficiently couples to this field. In particular, the induced magnetic field should efficiently drive flux through the inductor, ringing up the circuit. At the same time, the signal from the LC circuit should efficiently couple into the readout circuit. The LC circuit, when coupled to the readout circuit, should achieve a quality factor (Q) of one million. The virial velocity is $10^{-3}$ so the energy spread of a dark photon signal is $10^{-6}$. Thus, the signal power will increase linearly with Q up to $10^{6}$, and then become constant for higher quality factors. The LC circuit should be tunable and the resolution on the tuning should be better than one part in Q. The circuit geometries must cover nine orders of magnitude in frequency space from 100 Hz to 700 GHz. Additionally, in order to detect any signal from the dark photons, the LC circuit should lie in a shield that effectively blocks external radiation.

We design the straw-man experiment to operate at temperatures as low as 100 mK (although an initial demonstration experiment will be operated at 4 K). Temperatures below 100 mK may also be useful at the lower frequencies to further improve the sensitivity of the search, but they are not considered in this ``straw man'' analysis. The readout circuit is designed so that, when properly coupled, the thermal noise in the resonator at 100 mK can be resolved at frequencies below 1 GHz, and the quantum noise in both the resonator and amplifier limit the measurement above 1 GHz. In addition, the backaction from the readout circuit should not degrade the quality factor of the LC resonator below the target Q of one million.

\subsection{General Features of the Experiment}

Here we describe more general features of the straw-man design. A superconducting shield encasing the LC detection circuit is used to block external AC signals that could degrade the noise performance or be picked up as a signal in the resonator. The LC circuit is made of superconducting materials so that loss in the metals does not degrade the quality factor of the resonator.

The SQUID (Superconducting QUantum Interference Device) is an excellent magnetometer for signals below $\sim$ 1 GHz and is thus a good candidate for reading out the dark-photon-induced magnetic field. A convenient SQUID input coil inductance is in the range of hundreds of nanohenry (nH). The flux coupled into the SQUID is maximized for a pickup coil of equal inductance. If we were to use the resonant inductor as the pickup coil, the resonant capacitor would need to be undesirably geometrically large in the low-frequency kHz regime, and undesirably small in the high-frequency 100 MHz-700 GHz regime, in which case parasitics would dominate. Thus, the resonant inductor and pickup coil are separated, with a flux signal in the resonator coupling efficiently into the pickup coil. In the event that the SQUID shunt resistance degrades the cavity Q to below one million, we can shunt the input coil with an inductor to reduce the coupling and increase the Q.

Above 1 GHz, instead of using a SQUID, we read out the induced magnetic field using a quantum-limited parametric amplifier. In this case, the quantum noise of the amplifier and the resonator (48 mK/GHz for a phase-insensitive amplifier and 24 mK/GHz for a phase-sensitive amplifier)
dominates the measurement noise. \cite{caves} No transformer coil is required in this case.

\subsection{Implementation of the Strawman, Part 1: 100 Hz--1 GHz}

\begin{figure}
\centerline{\includegraphics[width=12cm]{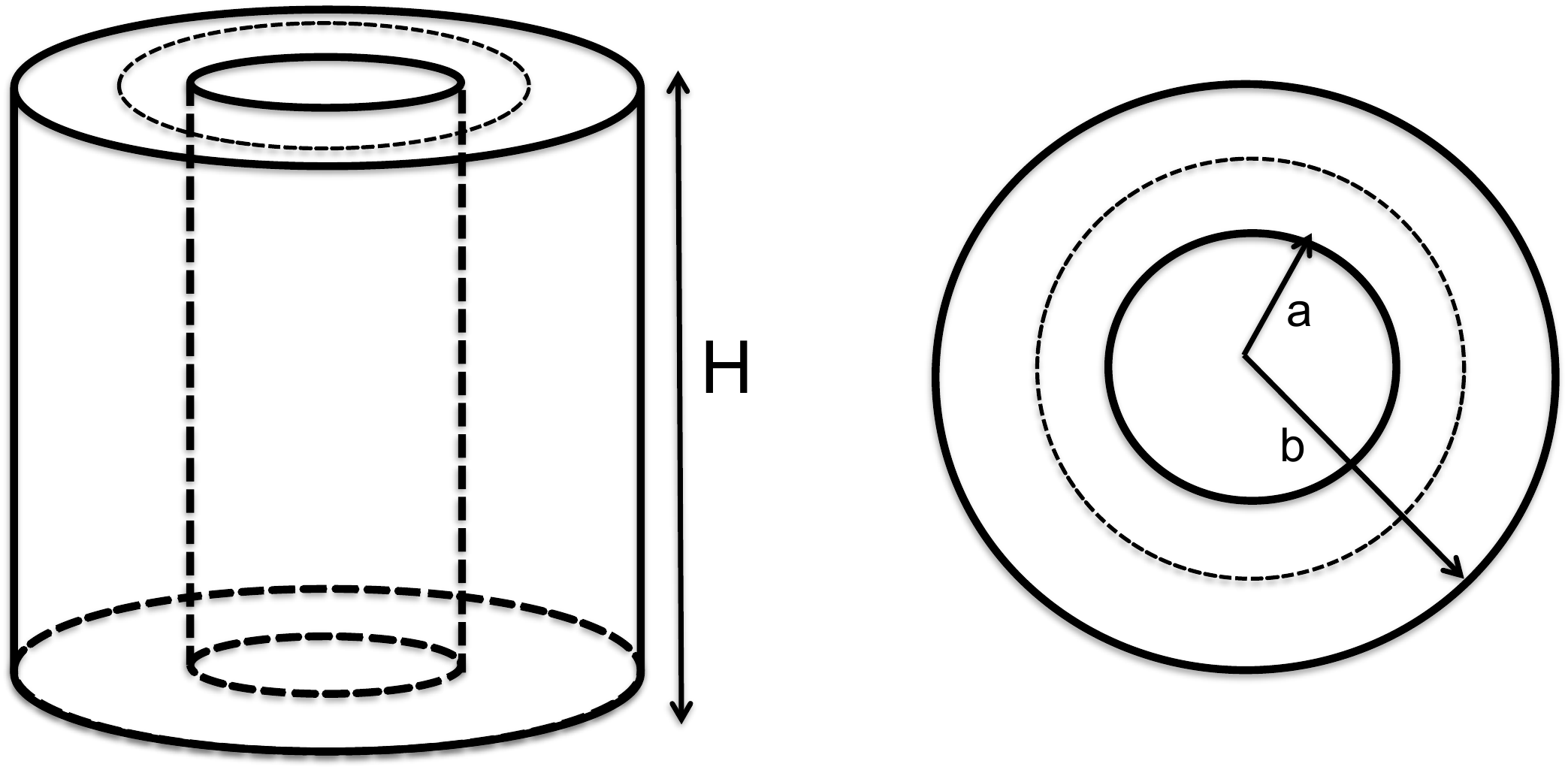}}
\caption{A hollow cylindrical slitted sheath that acts as a one-turn transformer coil. Left: Side view. Right: Top view. The inner radius of the sheath is a, the outer radius b, and the height H. All surfaces are solid except the top and bottom of the cylindrical hole in the middle and the dashed line on the upper surface which represents a slit. The coils of wire for the inductor in the hidden photon resonant detector can be wrapped as a toroidal solenoid around this sheath through the hole in the middle. The input coil for the SQUID can be attached on opposing sides of the slit. The superconducting shield encasing the entire experiment is not shown.} \label{fig:sheath}
\end{figure}

\subsubsection{ A Tunable LC Circuit}

A slitted sheath (see Fig. \ref{fig:sheath}) of inner radius $a$, outer radius $b$, and height $H$, has approximately an inductance of

\begin{equation}
L=\frac{\mu_{0}}{2\pi}H\mathrm{ln}\left( \frac{b}{a} \right)
\end{equation}

We have assumed that the width of the slit in the sheath is much smaller than the separation of inner and outer surfaces $b-a$. Thus, a sheath of inner radius of order $\sim$0.1 meters, outer radius order $\sim$ 1 meter, and height order $\sim$ 1 meter naturally has an inductance of a few hundred nH, which is appropriate for matching to the SQUID input coil. Thus, the sheath also functions as the SQUID pickup coil.

To strongly couple flux from the resonator into the sheath, we wrap the resonant inductor coil as a toroidal solenoid around the sheath through the hole in the middle. Turns of the coil can be wrapped in series or in parallel, depending on the desired LC resonator inductance.

Very rough tuning of the LC circuit frequency is conducted by varying the inductance of the inductor (by using a different number of turns, or adjusting series- and parallel-wiring), and by gross adjustments of the resonator capacitance.
To fine tune the circuit resonant frequency, we utilize a cryogenically tunable capacitor. The capacitance between two conducting surfaces is varied by changing the level of overlap between the surfaces and / or by adjusting the position of an insertable, low-loss dielectric such as high-purity sapphire. For a Q of one million, and capacitor overlap surface of roughly one square meter in area, the required resolution on plate adjustment is roughly one micron, which is achievable using a commercial piezoelectric. Both rough- and fine-tuning piezoelectrics may be implemented.

We note that both at low frequencies ($\simlt$1 kHz) and high frequencies ($\simgt$ 100 MHz), parasitic resonances are a concern. At low frequencies, a large inductor with $\sim$10000 turns of coil wrapped over $\sim$1 meter circumference is required. Not only is constructing this many turns challenging, but also, because of the high density of turns, there is significant parasitic capacitance between coil turns. This will result in parasitic resonances below the intended resonance frequency of the lumped element resonator, capacitively shunting some of the hidden photon signal. At high frequencies, degradation may also occur because of resonances of the sheath, shunting hidden photon signals at wavelengths comparable to and shorter than the dimensions of the sheath. For example, a one-meter tall sheath is expected to have a resonance at $\sim$ c/(2 m)=150 MHz, which will complicate detection of higher-frequency hidden photon signals. We can avoid this problem by having a few detection circuits, e.g. instead of one detection circuit with a meter-tall sheath, four detection circuits with 25 cm tall sheaths. As long as the circuits are within a coherence length of each other, we would not lose sensitivity by splitting the circuit. Additionally, the smaller detection circuits can all be put in the same superconducting shield.

The frequency range over which we can scan, given a fixed set of inductor coils, is limited by parasitics that dominate when the overlap between capacitor surfaces is small. It is likely that it will be necessary to change the coil set every decade in frequency.

To achieve the target Q of $10^{6}$, both the LC circuit and the resonator sheath can be made from high-purity niobium or niobium-titanium, which are used extensively in the construction of high-Q cavities. \cite{padamseerf} We will operate the experiment at well below the critical temperature to avoid thermally generated quasiparticles that would degrade the quality factor.

\subsubsection{Readout Architectures}

Here, we will discuss two readout achitectures: i) a dc SQUID and ii) a reactive ac SQUID coupled to a microwave resonator.

\paragraph{dc SQUIDS}\mbox{}\\

The dc SQUID is an interferometer based on two Josephson junctions that detects small magnetic fields; over the past five decades, it has become the standard tool in the measurement of such fields. \cite{squidhb}

\begin{figure}
\centerline{\includegraphics[width=12cm]{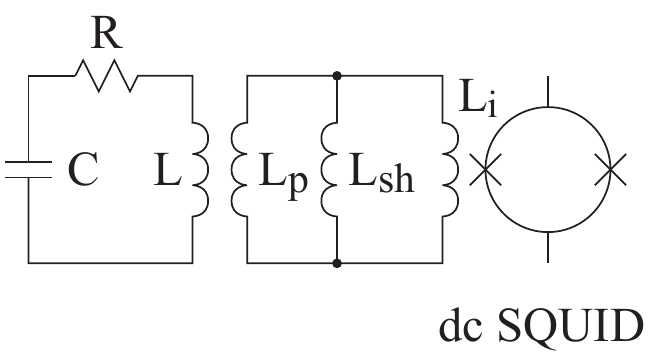}}
\caption{The dc SQUID readout architecture. The hidden photon resonant circuit (left) is flux-coupled to a transformer (middle) through the pickup coil, which has inductance $L_{p}$. The transformer, in turn, couples flux into a dc SQUID (right) through the input coil, which has inductance $L_{i}$. The variable inductive shunt $L_{sh}$ allows us to control the amount of back action from the SQUID, ensuring Qs of a million over a wide range of frequencies. The shunt inductor also allows us to set the level of cavity thermal noise coupling into the SQUID.} \label{fig:dcsquid1}
\end{figure}

We propose to use a dc SQUID to read out the dark photon signal in the 100 Hz- 10 MHz range, as shown in Fig. \ref{fig:dcsquid1}. In a dc SQUID-coupled LC circuit, the coupled Q is dominated by dissipation through the resistive shunts of the junctions. As we show in Appendix \ref{app:dcsquidapp}, despite this dissipation, for frequencies 100 Hz- 10 MHz, a loaded Q of at least one million can be achieved while at the same time, resolving the thermal noise of the cavity over a wide range of couplings and for temperatures down to 100 mK with typical commercial dc SQUID noise (a few $\mu \Phi_{0}/\sqrt{Hz}$ for commercial SQUIDs). \cite{muckdcsquid} This is facilitated by inserting a shunt inductor $L_{sh}$ across the input coil to reduce coupling to the SQUID and back-action from the SQUID to the resonator. 

Above 10 MHz, although loaded Qs of one million may be maintained by reducing the shunt inductance or the mutual inductances, the intrinsic flux noise of the SQUID, now comparable to the cavity thermal noise, will degrade the signal-to-noise ratio of the measurement. Novel transformer architectures may allow us to extend dc SQUID operation to 100 MHz. However, operation at frequencies above 100 MHz is difficult due to parasitic capacitance between the input coil and SQUID washer.\mbox{}\\

\paragraph{reactive ac SQUIDs}\mbox{}\\

Above 10 MHz, we propose to use a dissipationless, reactive ac SQUID coupled to a lithographed GHz-band resonator for the readout. \cite{hansma} \cite{likharevdynamics} This type of SQUID is now used in the readout of superconducting transition-edge sensors. \cite{matesphd} \cite{matesapl}

Reactive ac SQUIDs consist of a superconducting loop interrupted by one unshunted Josephson junction, rather than two shunted junctions, as in a dc SQUID. Consequently, the ac SQUID acts as a flux-dependent variable inductor (See Appendix \ref{app:acsquidapp}.) Because the dissipationless reactive ac SQUID does not have resistive shunts across the junction, the loss in the ac SQUID is dominated by sub-gap loss in the junction, and dielectric loss, and is significantly lower than the loss in the shunts of a dc SQUID. Consequently, we may couple the cavity more strongly to an ac SQUID than to a dc SQUID while still achieving an overall Q of one million.

\begin{figure}
\centerline{\includegraphics[width=12cm]{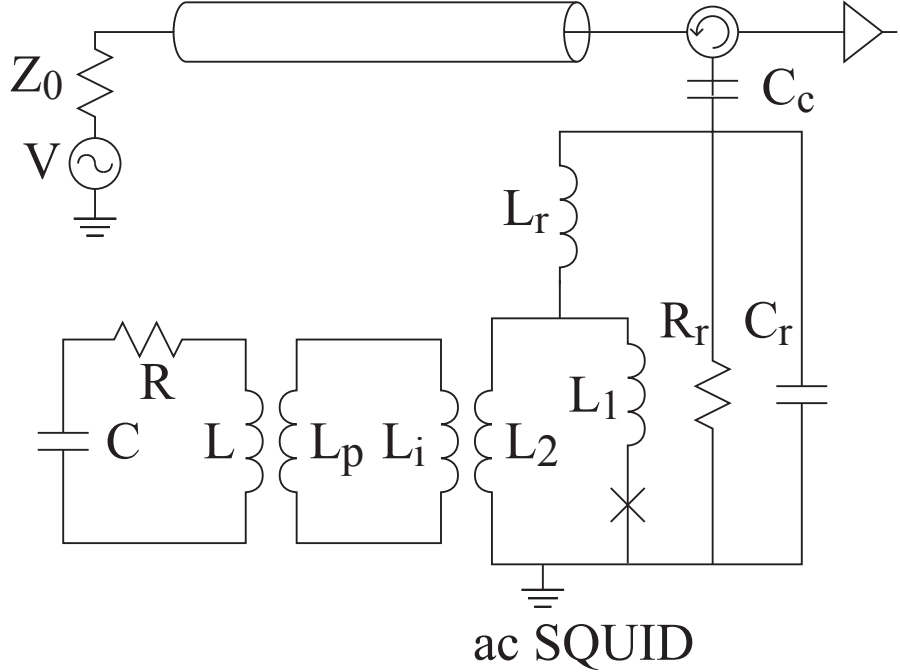}}
\caption{The ac SQUID readout architecture. The flux from the hidden photon resonant circuit is coupled into a reactive ac SQUID, in a similar manner as for the dc SQUID in Fig. \ref{fig:dcsquid1} (with the exclusion of the inductive shunt). The microwave resonator coupled to the ac SQUID (directly below the circulator), denoted by subscript $r$, is driven by source represented by $V$ (to the left of the circulator). The flux modulates the phase of the reflected drive tone. This change in phase can be read out by a low-noise amplifier, e.g. a HEMT or quantum-limited parametric amplifier (to the right of the circulator).} \label{fig:acsquid1}
\end{figure}

The ac SQUID is coupled to a lithographic microwave resonator that is attached to a superconducting lithographic feedline, as shown in Fig. \ref{fig:acsquid1}. The lithographic ac SQUID resonator is driven at close to its resonant frequency, typically in the 5-10 GHz frequency range. The signal from the dark-photon LC resonator, which operates at  $\simlt$ 1 GHz, is coupled into the ac SQUID as an ac magnetic flux signal, causing a change in its Josephson inductance. This in turn causes a change the resonance frequency of the ac SQUID resonator, which results in a periodic phase shift in the reflected microwave signal on the feedline, and is measured as sidebands in the reflected microwave signal. This phase modulation is read out using a low-noise commercial microwave amplifier, such as a cryogenic HEMT, or quantum-limited parametric amplifier.

If there is a signal in the dark-photon LC resonator, the resulting periodic phase shift in the reflected microwave signal on the feedline will result in a frequency component at the ac SQUID drive frequency, and two accompanying sidebands separated by the dark photon frequency. So that sensitivity is maximized, we require that these sidebands are well within the bandwidth of the feedline resonator. We also require that the flux modulation not move the resonance frequency more than one-half of a bandwidth. In the appendix, we show that for reasonable circuit geometries/parameters, these requirements are readily met. We also show that over the entire frequency range 10 MHz-1 GHz, the sub-gap junction loss and feedline impedance should not degrade the quality factor below one million and that the thermal noise dominates over the intrinsic SQUID flux noise and amplifier noise noise down to 100 mK--that is, the thermal noise is the dominant noise contribution in the circuit. (Note that the ac SQUID technique may be used to probe the 100 Hz-10 MHz regime, though the readout may be more complicated than that using a dc SQUID.) However, above 1 GHz, the quantum noise of the resonator and amplifier dominates thermal noise. This motivates the direct parametric amplifier readout scheme present in PART II.

\subsection{Implementation of the Strawman, Part 2: 1 GHz--700 GHz}

\subsubsection{Tunable LC Circuit}

Above 1 GHz, instead of utilizing an inductive sheath as a transformer coil, we can use it as the inductor of the resonant circuit. As we discussed in PART I, a large pickup sheath that efficiently couples to the hidden photon field will have resonances lower than the hidden photon frequency that we are probing; we can avoid these resonances by using many circuits within a coherence length. However, as we go to higher frequencies to avoid resonances, we require more detection circuits, increasing readout complexity. As such, a total volume of 1 cubic meter, as in PART I, is more difficult to achieve without a highly multiplexed readout circuit. Regardless, we estimate that for this higher frequency regime, the total detector volume will be at minimum several times the cube of the wavelength.

An alternative is to use a cavity instead of a lumped-element resonator. We note that above 700 GHz, which is approximately the gap frequency of niobium, the cavity becomes unacceptably lossy (low Q). Other superconductors, such as NbTiN (gap frequency of 1.1 THz) may allow extension of this experiment into higher frequency ranges.

To tune each of the LC circuits, we may utilize piezoelectric tuning. The capacitor will be approximately 1 cm x 1 cm, so we require 10 nm spatial resolution for one part in a million tuning, which is achievable with some piezoelectrics.

Given the need to replace detection circuits as we approach sheath resonances, we estimate that we will be able to scan with the same setup a factor of two in frequency.

\begin{figure}
\centerline{\includegraphics[width=12cm]{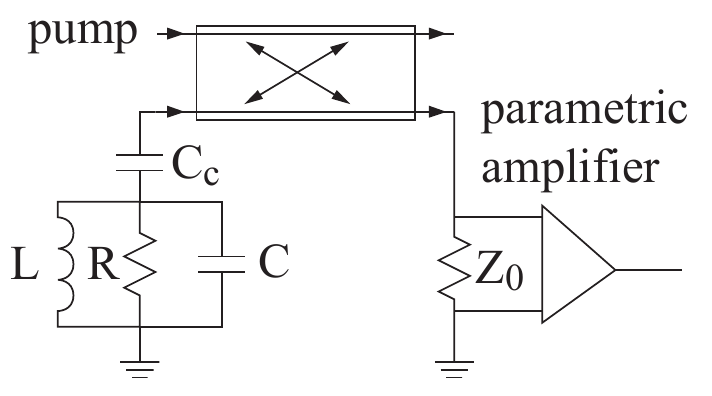}}
\caption{A potential parametric amplifier readout architecture. The hidden photon readout circuit is capacitively coupled to a feedline and amplified using a quantum-limited parametric amplifier. A pump power supply for the amplifier is fed through a directional coupler.} \label{fig:qlamp1}
\end{figure}

\subsubsection{Readout architecture}
The resonator will be read out using a quantum-limited parametric amplifier. \cite{lehnertjpa} This type of readout has been proposed before for the Axion Dark Matter Experiment. \cite{admxsqueeze} One option is to capacitively couple the resonator to a feedline, which is connected to the amplifier. The readout circuit is displayed above in Fig. \ref{fig:qlamp1}. Since we are well below the gap frequency and will operate well below the critical temperature, the quality factor is dominated by the coupling to the feedline (i.e. dissipation in the real characteristic impedance of the feedline and the termination resistance $Z_0$.) The resonator frequency and coupling Q can be found in a manner similar to that established in Appendix B:
\begin{equation}\label{eq:omegarGHz}
\omega_{r}=\frac{1}{\sqrt{L(C+C_{c})}}
\end{equation}
and
\begin{equation}\label{eq:QcGHz}
Q_{c}=\frac{1}{\omega_{r}C_{c}Z_{0}\omega_{r}^{2}LC_{c}}
\end{equation}

For the frequency range 1 GHz-700 GHz, the quantum noise temperature (48 mK/GHz for a phase-insensitive amplifier, and 24 mK/GHz for a phase-sensitive amplifier) will be larger than the operating temperature of the experiment, nominally ~100 mK. In this regime, we will be dominated by quantum noise rather than thermal noise.

Another option under investigation is read out using a single-photon-counting (SPC) detector in the form of a superconducting quantum bit architecture. Such schemes have been proposed for the Axion Dark Matter Experiment (ADMX) and are under investigation. \cite{admxspcamp} A single-photon detector would evade the quantum limit and allow much lower noise powers, but have limited dynamic range, which may make it more difficult to understand and eliminate interfering signals.

Another option in the readout, which has also been proposed for ADMX, is squeezing the thermal noise using a phase-sensitive parametric amplifier. This, in the SPC scheme, has the potential to reduce the overall noise by another order of magnitude. \cite{admxsqueeze}

\section{Sensitivity estimate}
\label{sec:sensitivity}

From Eq.~\ref{eq:B-signal}, the signal power is given by
\begin{align}
P_{\rm sig} &= \frac{\omega}{Q} \times \text{(stored energy)}
\approx \frac{\mhidden}{Q} B_{\rm sig}^2 V_{\rm ind}
\\
&\approx  Q \, \varepsilon^2 \, m_{\gamma'} \rho_{\rm DM} V_{\rm ind} \times
\begin{cases}
\nu^2 V_{\rm ind}^{2/3} & \quad \nu \simlt V_{\rm ind}^{-1/3}
\\
1 & \quad \nu \simgt V_{\rm ind}^{-1/3} \, .
\end{cases}
\end{align}
where again $V_{\rm ind}$ is the volume of the inductor,
and the lower-frequency scaling accounts for the suppression of the observable B-field inside the shielding.
We assume a typical volume of $1\,{\rm m}^3$. However, as discussed above, for high frequencies the volume of the inductor is limited by $1/\nu^3$ rather than by practical size constraints. We therefore assume
\begin{equation}
V_{\rm ind} \approx \begin{cases}
1\, \rm m^3  & \quad \nu \simlt V_{\rm ind}^{-1/3}
\\
1/\nu^3 & \quad \nu \simgt V_{\rm ind}^{-1/3}
\end{cases}
\qquad\qquad \text{(basic setup)} \, .
\end{equation}
for our canonical setup.
A larger total volume can be reached by multiplexing many inductors within a single coherence length of the hidden-photon field, allowing the full m$^3$ to be maintained in principle up to frequencies as high as $\nu \sim 10^3 /$m. We therefore also consider a multiplexed setup, where
\begin{equation}
V_{\rm ind} \approx \begin{cases}
1\, \rm m^3  & \quad \nu \simlt 10^3 V_{\rm ind}^{-1/3}
\\
10^9/\nu^3 & \quad \nu \simgt 10^3 V_{\rm ind}^{-1/3}
\end{cases}
\qquad\qquad \text{(multiplexed setup)} \, .
\end{equation}

At low frequencies the integration time does not allow use of the full $Q$-factor of $10^6$. Take each frequency step to be scanned for a time $t_{\rm step} \approx t_\text{e-fold}/Q$, where $t_\text{e-fold}$ is the time taken to scan one e-fold in frequency. We would like $t_{\rm step}$ to be at least as large as $Q/\nu$ to allow the full signal to ring up. Therefore at the lowest frequencies we will allow the resonator to have a lower $Q$-factor, and take
\begin{equation}
Q \approx \begin{cases}
10^6 & \quad \nu \simgt 10^{12}/t_\text{e-fold}
\\
\sqrt{\nu \, t_\text{e-fold}} & \quad \nu \simlt 10^{12}/t_\text{e-fold} \,\, .
\end{cases}
\end{equation}

The noise power is given by
\begin{gather}
P_{\rm noise} = \frac{\delta \nu \, T_{\rm eff}}{\sqrt{\delta \nu \, t_{\rm step}}} \approx \sqrt \frac{\nu}{t_\text{e-fold}} T_{\rm eff} \, ,
\end{gather}
where $\delta \nu \approx \nu/Q$ is the bandwidth, and we have accounted for quantum noise by using an effective noise temperature of
\begin{equation}
T_{\rm eff} = (n_\omega + 1) \omega = \frac{\omega\, e^\frac{\omega}{T}}{e^\frac{\omega}{T} -1}
\approx \begin{cases}
T & \quad \mhidden < T
\\
\mhidden & \quad \mhidden > T \,
\end{cases}
\end{equation}
(N.B. we are using  units where $\hbar = k_B = c = 1$).

We estimate the reach in $\varepsilon$ by requiring a signal-to-noise ratio $P_{\rm sig}/P_{\rm noise} \simgt 1$. Combining the above equations gives
\begin{align}
\varepsilon\sim&
10^{-17} \times
\sqrt{\frac{10^6}{Q_{\rm max}}}
\bigg( \frac{10^{-6} \, \eV}{\mhidden} \bigg)^{1/4}
\bigg( \frac{30 \, \rm days}{t_\text{e-fold}} \bigg)^{1/4}
\sqrt{\frac{T}{0.1 \, \rm K}}
\sqrt{\frac{0.3 \, \GeV/ \cm^3}{\rho_{\rm DM}}}
\sqrt{\frac{1\, \rm m^3}{V}}
\nonumber \\
&\qquad \times \max \bigg[ \bigg(\frac{2\pi \, Q_{\rm max}^2}{m_{\gamma'} t_\text{e-fold}} \bigg)^{1/4}, 1 \bigg]
\times \max \bigg[ 1, \sqrt{\frac{\mhidden}{T}} \bigg]
\times \max\bigg[\frac{2\pi}{m_{\gamma'} V^{1/3}} , \bigg(\frac{m_{\gamma'}V^{1/3}}{2\pi}\bigg)^{3/2} \bigg] \, ,
\label{eq:epsilon-reach}
\end{align}
where $V\!\sim\! 1 \, \rm m^3$ is the volume of the inductor at lower frequencies, $Q_{\rm max} \approx 10^6$ is the full $Q$-factor, and we are assuming only a single (smaller) resonator in the high frequency regime.
In this regime, the improvement achievable by multiplexing many detectors can be described by making the replacement
\begin{equation}
\max\bigg[\frac{2\pi}{m_{\gamma'} V^{1/3}} , \bigg(\frac{m_{\gamma'}V^{1/3}}{2\pi}\bigg)^{3/2} \bigg]
\longrightarrow
\max\bigg[\frac{2\pi}{m_{\gamma'} V^{1/3}} , 1,  \bigg(\frac{m_{\gamma'}V^{1/3} v_{\rm DM} }{2\pi}\bigg)^{3/2} \bigg]
\end{equation}

We consider setups operating in three frequency regimes: a ``low-f'' regime from $\sim\!100 \, \Hz$--$10 \, \MHz$, a ``mid-f'' regime from $\sim\!10 \, \MHz$--$1 \, \GHz$, and a ``high-f'' regime from $\sim\! 1 \, \GHz$--$700 \, \GHz$. These correspond to the regimes in which the signal is read out using an ac SQUID, a dc SQUID, and a parametric amplifier, as described above. In all regimes we assume the experimental parameters indicated in Eq.~\ref{eq:epsilon-reach}, except that below 100\,kHz we assume $t_\text{e-fold} = 90 \,$days to allow a longer ring-up time. 
We plot our estimated reach in figure~\ref{fig:sensitivity}, along with various other constraints compiled from~\cite{Jaeckel:2010ni, Arias:2012az, An:2013yfc, Graham:2014sha}.
The solid blue regions show our estimated reach in the three frequency ranges.
The dashed blue line shows the improvement achievable in principle by multiplexing at high frequencies, while maintaining the $1 \, \rm m^3$ maximum volume.

\section{Conclusions}
\label{sec:conclusions}

Hidden photons arise naturally in many theories of particle physics beyond the Standard Model, and are a viable dark matter candidate. 
As a dimension 4 operator, kinetic mixing is the dominant way in which the standard model can interact with a hidden photon, since these effects can be generated at a high scale without their observable effects being suppressed by that scale. 
There is thus strong motivation to search for such particles. However, in light of our ignorance of the ultraviolet structures of particle physics, we do not know the masses or coupling strengths of these particles. 

Experimental searches for hidden photons can make use of certain of their properties: their coupling to charged SM particles, their ability to penetrate shielding (due to the smallness of this coupling), and, if they make up the dark matter, the nearly-monochromatic energy spectrum of their local abundance. 
Currently, there are three classes of experiments that can search for hidden photons beyond astrophysical limits. 
One class ~\cite{Essig:2009nc, Batell:2009di} uses beam dumps in particle colliders to probe (non dark-matter) hidden photons in the MeV--GeV mass range. 
A second class is high-Q microwave cavity experiments that produce longitudinal hidden photons and detect them behind a shield in a ``Light Shining Through a Wall" experiment. 
These currently have only limited sensitivity in a frequency range around GHz, although they appear to have the potential to probe new parameter space over a much wide range of masses. 
The third class uses resonant electromagnetic detectors to search for dark-matter hidden photons. The only current example is the ADMX~\cite{Asztalos:2011bm} experiment, which is sensitive to dark-matter hidden photons in the 0.3--3 GHz~\cite{Arias:2012az} mass range. 

No current experiment is sensitive to hidden photons beyond these limited ranges. 
However, hidden photons can exist over a much wider range of masses, and can make up the dark matter of the universe as long as their mass is greater than $\sim\! 10^{-22}~\eV$, corresponding to frequencies above $\sim\!10^{-8}$~Hz.  
In light of the strong case for their existence, a variety of experimental approaches must be developed to cover the full parameter space. 
 
In this paper, we proposed the use of tunable, high-$Q$ LC resonators to search for hidden photon dark matter over a wide frequency range. Previous work \cite{Nelson:2011sf, Arias:2012az} has largely focussed on placing bounds on such particles from current experiments. By utilizing the coherent nature of the dark matter signal and the existence of precision measurement protocols to detect weak electromagnetic signals in this frequency range in an optimized experiment, we show that it is feasible to probe hidden photon dark matter well past present bounds, both in frequency and coupling. The ability to probe deep into this parameter space exists despite a suppression of the signal in the presence of shielding, which causes the hidden-photon field to rotate into the sterile basis at leading order. 

We have presented preliminary ``straw man'' designs for cubic-meter-scale experiments covering the frequency range $\sim\! 100 \, \Hz$--$700 \, \GHz$.
In the future, lower limits may be achievable by cooling the experiment in a dilution refrigerator to lower temperatures (perhaps 10 mK), although this will also lower the frequency at which quantum noise dominates.

At present, the lower limit on our frequency range $\sim$ kHz is set by the technological limitations of obtaining low frequency, high-$Q$ LC resonators. It might be possible to probe hidden photons below this frequency range through experiments such as CASPEr \cite{Graham:2013gfa, Budker:2013hfa} that utilize precision magnetometry coupled with NMR. This possibility arises because the hidden-photon dark matter creates a real magnetic field inside a shielded region. This magnetic field can potentially be observed using a suitably designed NMR device. It may also be possible to use devices such as ion clocks to search for the electric field induced by the hidden photon inside a shielded region. This might be an attractive avenue to pursue ultra low frequency hidden photons. We leave the development of these ideas for future work.  

Much like searches for axion dark matter \cite{Asztalos:2011bm, Graham:2013gfa, Budker:2013hfa}, a positive signal in this experiment can be verified in many ways. In this frequency range, the dark matter signal is coherent over macroscopic distances ($\simgt$ km) and hence two independent detectors constructed within this distance should have signals that are correlated in frequency, phase and direction over the course of the experiment. 
The vector nature of the dark matter will also result in distinctive correlations between differently oriented detectors.
In fact, at lower frequencies where the signal is coherent over long periods of time, it might be possible to observe the relative rotation between the earth and the local direction of the hidden photon dark matter's electric field. In the proposed resonant scheme, the experiment only requires a few seconds to search for dark matter in any particular frequency band. If the experiment detects a positive signal in that band during this short time, the device can operate at that band for a longer time to see if the signal builds in a manner consistent with a dark matter signal. Furthermore, a confirmed dark matter signal in such an experiment would lead not just to the discovery of dark matter, but through the measurement of the coherence properties of the signal also lead to a probe of the local velocity structure of dark matter. 

Over the past two decades, WIMP direct detection experiments have made tremendous progress, with regular increases in their sensitivity. These advances were made possible due to the scalable nature of these experiments wherein a small, well studied, initial apparatus can be scaled to larger sizes. Similar approaches seem possible in the search for these ultra light particles such as axions and massive vector bosons. Since these  are also prime dark matter candidates, it is important to develop techniques to search for them. In fact, the identification of the properties of dark matter may offer a unique way to probe high energy physics. The discovery of WIMPs would be a portal into new physics at the TeV scale. In contrast, since particles such as axions and massive vector bosons can arise from ultra-high energy physics, close to the scales of unification and quantum gravity, their discovery would open one of the very few ways in which we can glimpse these fundamental scales.

\section*{Note Added}

As this work was being completed Ref.~\cite{Arias:2014ela} appeared, which also proposes the use of LC resonators to search for hidden-photon dark matter in a section of the frequency range we consider. 
The two papers agree in their central concept, 
but the present work also goes beyond the work of Ref.~\cite{Arias:2014ela} in several key aspects:
1) We consider shielding and its important (but subtle) effect on the signal.
2) We present a careful and systematic approach to calculating the signal. 
3) We have presented the essential features of a realistic experimental design, including the pickup circuit itself, the shielding, and the readout architecture. 
4) While the sensitivity estimated in Ref.~\cite{Arias:2014ela} is limited by the magnetometer itself, we have also taken into consideration thermal and quantum noise, and find them to be more limiting than magnetometer noise.
Regarding the first two points, we feel that Ref.~\cite{Arias:2014ela} is somewhat confusing or even misleading to the reader, and that some clarification is required, which we attempt below. 

A naive estimate of the observable fields, based on a single test-charge far away from any conductors, would suggest an observable $E$-field of size $\varepsilon E'$,
and a much smaller $B$-field of size $B\sim\! v \times \varepsilon E'$, where $v\sim10^{-3}$. 
However, the presence of conductors, including shielding, affects the observable fields.
Since electromagnetic shielding is an essential part of a realistic experimental setup, we include it as a core ingredient of our calculation. 
As we discuss, shielding fixes the external boundary conditions required to solve the modified Maxwell equations. 
The result is a parametrically suppressed observable $E$-field of size $E \!\sim\! \varepsilon (m R)^2 E'$, and an observable $B$-field of size $B \!\sim\! \varepsilon (m R) E'$, which is thus the dominant signal field (in the $m R \ll 1$ regime).

While Ref.~\cite{Arias:2014ela} arrives at the same parametric result as us for the signal $B$-field, it does so without considering shielding, without a systematic calculation, and without a mentioning the observable field $E$-field or why it is suppressed compared with the naive estimate. 
The calculation in Ref.~\cite{Arias:2014ela} is very brief, and is based on a similar calculation for an axion search presented in Ref.~\cite{Sikivie:2013laa}. 
Neither calculation mentions the physical boundary conditions for the modified Maxwell equations, although without these the solution cannot formally even be determined. 
This is perhaps acceptable in the axion-detection calculation of Ref.~\cite{Sikivie:2013laa}, since in that case the source term is spatially confined (it only exists where a static magnetic field is applied). 
This means that the signal fields in the proposed axion experiment would be parametrically the same whether shielding was used, or the experiment was unshielded and placed far from other conductors. 
However, this is not true of the hidden-photon case considered here and in Ref.~\cite{Arias:2014ela}, where the source term added to Maxwell's equations extends over the whole of space. 
This difference makes it inappropriate to use the calculation of Ref.~\cite{Sikivie:2013laa} in the hidden-photon context without specifying the presence of a shield with conducting boundary conditions.

To see the physical importance of this, consider an \emph{unshielded} hidden-photon experiment set up far from other conductors. 
In that case, we find that a $B$-field of size $B \!\sim\! \varepsilon (m R) E'$ would still be generated in the inductor, but that there would also be an \emph{unsuppressed} observable $E$-field of size $E\!\sim\! \varepsilon E'$. 
This would drive an appropriately positioned capacitor, giving a much larger signal than that picked up by the inductor. 
While such a setup is of course impractical, it demonstrates the importance of shielding and the subtlety of the calculation.

\section*{Acknowledgements}
We would like to thank C.L. Kuo, L. Page, S. Thomas, and A. Tyson
for many useful discussions. PWG acknowledges the
support of NSF grant PHY-1316706, DOE Early Career Award
DE-SC0012012, the Hellman Faculty Scholars program, and the Terman
Fellowship. JM is supported by grant DE-SC0012012. SR acknowledges the support of NSF
grant PHY-1417295. YZ is supported by ERC grant BSMOXFORD no.
228169. SC acknowledges the support of a NASA NSTRF Fellowship.

\appendix

\section{Observable fields inside a conducting shield}
\label{sec:signal-in-shield-full-calc}

In this appendix, we determine the observable $E$ and $B$-fields generated inside a conducting shield in the presence of the dark matter hidden-photon field $A'_\mu(\vec x, t)$, in the limit that the shield is much smaller than the inverse frequency. These are the fields that will drive the LC circuit placed inside the shield in low-frequency and mid-frequency setups proposed above.
The theory governing the behavior of $E$ and $B$-fields in the presence of a background hidden-photon field is described in detail in Ref.~\cite{Graham:2014sha}. We follow closely the formulation and results presented there.

Since the oscillating hidden-photon field corresponds to an $E'$-field of size $E'\approx m_{\gamma'}A' \approx \sqrt{\rho_{\rm DM}}$, which couples with strength $\varepsilon$ to electric charges, one might expect the observed field to be an $E$-field with size $E_{obs} \sim \varepsilon \sqrt{\rho_{\rm DM}}$.
However, the presence of the conducting shield has a significant effect.
As appropriate for this setup, we take the period of oscillation to be much longer than the length scale of the shield (i.e. $m_{\gamma'} R \ll 1$).
As we shall see, the dominant observable field is in fact a $B$-field, with size suppressed by a factor $\varepsilon m_{\gamma'} R$ compared to $E'$.

\vspace{4pt} \paragraph{\bf Equivalent problem in EM}

Consider a single (complexified) mode of the hidden photon field, $A'_\mu \!\propto\! e^{i (\omega t - \vec k \cdot \vec x)}$.
Its effect in vacuum is identical to that of an oscillating current density~\cite{Graham:2014sha}, given by
\begin{equation}
\vec \jmath_{\rm eff}(\vec x, t) = - \frac{i \varepsilon}{\omega} \big(m_{\gamma'}^2 \vec E'(\vec x, t) + \vec k (\vec k \cdot \vec E'(\vec x, t))\big) \, ,
\label{eq:effective-current-1}
\end{equation}
where $\vec E' = i \vec k V' - i \omega \vec A'$ and $\omega V' = \vec k \cdot \vec A'$. Rewriting in terms of $\vec A'$ rather than $\vec E'$, this simplifies to
\begin{equation}
\vec \jmath_{\rm eff}(\vec x, t) = - \varepsilon m_{\gamma'}^2 \vec A'(\vec x, t) \, .
\label{eq:effective-current-2}
\end{equation}
(This holds for any $\vec A'(\vec x, t)$, not just a plane wave. Note that when working in the interaction basis, Eq.~\ref{eq:effective-current-2} can be read off directly from the Lagrangian term $\mathcal L \supset \varepsilon m_{\gamma'}^2 A_\mu A'^\mu$.)

The observed $E$ and $B$-fields within the shield are determined by Maxwells equations, with the conductor boundary condition $\vec {d S} \times \vec E=0$ on the surface of the shield, and in the presence of an effective current density (and corresponding charge density) given by Eq.~\ref{eq:effective-current-2}.

\vspace{4pt}\paragraph{\bf Example}

As an example, take the shield to be a hollow cylinder of radius $R$ (and of any length), aligned along the $z$-axis, and take the $\vec A'$ field to point along the same axis. Ignoring the small dark matter velocity, we have
\begin{align}
\vec \jmath_{\rm eff}(\vec x, t) &= - \varepsilon m_{\gamma'}^2 A' e^{i m_{\gamma'}t} \hat z \\
\varrho_{\rm eff}(\vec x, t) &= 0\, ,
\end{align}
where $A'$ is a constant.
The observable $E$ and $B$-fields then satisfy
\begin{align}
(\nabla^2 - \partial_t^2)\vec E_{obs} = \partial_t \vec \jmath_{\rm eff} + \vec \nabla \varrho_{\rm eff}
\quad &\longrightarrow \quad
(\nabla^2 + m_{\gamma'}^2)\vec E_{obs} = - i \varepsilon m_{\gamma'}^3 A' e^{i m_{\gamma'}t} \hat z \\
\vec \nabla \times \vec E_{obs} = - \partial_t \vec B_{obs}
\quad &\longrightarrow \quad
\vec B_{obs} = \frac{i}{m_{\gamma'}} \vec \nabla \times \vec E_{obs} \\
\hat z \cdot \vec E_{obs} \!&\,= 0 \quad \text{at }r =R \, .
\label{eq:B-in-cylinder}
\end{align}

The solution is
\begin{align}
\vec E_{obs} &= - i \varepsilon m_{\gamma'} A' e^{i m_{\gamma'}t} \Big(1-\frac{J_0(m_{\gamma'} r)}{J_0(m_{\gamma'} R)} \Big) \hat z
\approx i \varepsilon \sqrt{\rho_{\rm DM}} e^{i m_{\gamma'}t} \hat z \times m_{\gamma'}^2 (R^2 - r^2) \\
\vec B_{obs} &= - \varepsilon m_{\gamma'} A' e^{i m_{\gamma'}t} \frac{J_1(m_{\gamma'} r)}{J_0(m_{\gamma'} R)} \hat \phi
\approx -\varepsilon \sqrt{\rho_{\rm DM}}e^{i m_{\gamma'}t} \hat \phi \times m_{\gamma'} r \, ,
\end{align}
where we have used $\rho_{\rm DM} \approx m_{\gamma'}^2 {A'}^2 /2$, and taken $m_{\gamma'} R \ll 1$.
We see that the largest observable field is the $B$-field, which is generated by currents in the shield walls and points in the $\hat \phi$ direction.

\vspace{4pt}\paragraph{\bf General solution}

The conducting shield is essentially an electromagnetic cavity driven far off resonance.
This allows the fields inside it to be determined with the standard methods for driven cavities (see e.g. chapter 1 of~\cite{Hill}).
The boundary condition for the $E$-field allows it to be decomposed into a complete orthonormal basis of the form
\begin{align}
\vec E_{obs}(\vec r, t) = \Big( \sum_n c_n \vec E_n(\vec r) + \sum_p d_p \vec F_p(\vec r) \Big) e^{i \omega t}  \, .
\label{eq:E-mode-decomposition}
\end{align}
Here $\vec E_n(\vec x)$ are the vacuum cavity modes of the shield's interior, satisfying $\nabla^2 \vec E_n = -\omega_n^2 \vec E_n$ and $\vec \nabla \cdot \vec E_n = 0$. $\vec F_p(\vec x)$ are a set of irrotational functions satisfying $\vec F_p = - \vec \nabla \Phi_p$ and $\nabla^2 \Phi_p = -\tilde \omega_p^2 \Phi_p$.
They also satisfy the boundary conditions $\vec {d S} \times \vec E_n = 0$ and $\Phi_p = 0$ on the inner surface of the shield. Using $\vec \nabla \times \vec E = - i \omega \vec B$, we can write the interior $B$-field as
\begin{align}
\vec B_{obs}(\vec r, t) = \sum_n c_n \frac{i}{\omega} \vec \nabla \times \vec E_n(\vec r) \, e^{i \omega t} = \sum_n c_n \frac{\omega_n}{\omega} \vec B_n(\vec r) \, e^{i \omega t}\, ,
\label{eq:B-mode-decomposition}
\end{align}
where $\vec B_n$ are the $B$-fields of the vacuum cavity modes given by $\vec \nabla \times \vec E_n \equiv - i \omega_n \vec B_n$.
Note that since the cavity is driven far off resonance, we do not need to consider the build up of this signal over time, and can instead just take the infinite time solution. The coefficients are given by
\begin{align}
c_n &= \frac{-i \omega}{\omega_n^2 - \omega^2} \frac{ \int d^3 x \, \vec E_n^*(\vec x) \cdot \vec \jmath_{\rm eff}\, (\vec x, 0) }{ \int d^3 x \, |E_n(\vec x)|^2 }
\label{eq:coefficient-ODE-1}\\
d_p &= \frac{i}{\omega} \frac{ \int d^3 x \, \vec F_p^*(\vec x) \cdot \vec \jmath_{\rm eff}\, (\vec x, 0) }{ \int d^3 x \, |F_p (\vec x)|^2 } \, ,
\end{align}
where the integrals are taken over the interior of the shield.

\vspace{4pt}\paragraph{\bf Leading order terms}

Since the dark matter is non-relativistic and we are considering a shield of characteristic size $R \ll 1/m_{\gamma'}$, $\vec \jmath_{\rm eff}$ is approximately constant in space and can be written as
\begin{equation}
\vec \jmath_{\rm eff}(\vec x, t) = - \varepsilon m_{\gamma'}^2 A' \hat n \, e^{i m_{\gamma'} t}(1 - i m_{\gamma'} \vec v \cdot \vec x + \mathcal O (m_{\gamma'} R v)^2) \, ,
\end{equation}
where $A'$ and $\hat n$ are constants.
For the coefficients $c_n$, the constant term in $j_{\rm eff}$ dominates, giving
\begin{gather}
c_n \approx i \varepsilon (m_{\gamma'} R)^2  m_{\gamma'} A' \hat n \cdot \frac{ \int d^3 x \, \vec E_n^*(\vec x)}{\omega_n^2 R^2 \int d^3 x \, |E_n(\vec x)|^2 } \, ,
\end{gather}
where we have used $\omega_n \simgt 1/R \gg  \omega =  m_{\gamma'}$. The numerator is generally non-vanishing for a large number of modes.
However for the coefficients $d_p$, the numerator always vanishes if we keep only the constant term in $j_{\rm eff}$, because $\int d^3 x \, \vec F_p = - \oint \vec{d S} \, \Phi_p = 0$. We therefore need the term proportional to $m_{\gamma'} \vec v \cdot \vec x$, giving
\begin{gather}
d_p \approx - \varepsilon m_{\gamma'} A' \hat n \cdot \frac{ \int d^3 x \, (m_{\gamma'} \vec v \cdot \vec x) \vec F_p^*(\vec x) }{ \int d^3 x \, |F_p (\vec x)|^2 }  \, .
\end{gather}

Using $\rho_{\rm DM} = m_{\gamma'}^2 {A'}^2 /2$, we can now see the parametric size of the observed fields
\begin{align}
\vec E_{obs} &= \varepsilon \sqrt{\rho_{\rm DM}} \times \big( \mathcal O (m_{\gamma'} R)^2 + \mathcal O (m_{\gamma'} R \, v_{\rm DM}) \big) \\
\vec B_{obs} &= \mathcal \varepsilon \sqrt{\rho_{\rm DM}} \times O (m_{\gamma'} R)  \, .
\end{align}
We see that the largest observable field inside the shield is the $B$-field, whose size is suppressed by a factor $\varepsilon m_{\gamma'} R$ relative to the $E'$-field. The observable $E$-field is further suppressed by a factor of either $m_{\gamma'} R$ or $v_{\rm DM}$, whichever is larger. We can therefore neglect $E_{obs}$, and focus on detecting the $B$-field, which is given in full generality by
\begin{align}
\vec B_{obs}(\vec r, t) = i \varepsilon \, m_{\gamma'} R \, \sqrt{2 \rho_{\rm DM}} \hat n_i  \sum_n \frac{ \int d^3 x \, E_{n \, i}^*(\vec x)}{\omega_n R \int d^3 x \, |E_n(\vec x)|^2 } \vec B_n(\vec r) \, e^{i \omega t}\, .
\label{eq:B-full-solution}
\end{align}
where again, $\vec E_n(\vec x)$ and $\vec B_n(\vec x)$ are the cavity modes of the shield's interior.


\section{dc SQUIDs: Back-impedance and noise}
\label{app:dcsquidapp}

We demonstrate that, for temperatures down to 100 mK and for resonator frequencies below 10 MHz, loaded (i.e. coupled to dc SQUID) quality factors of one million can be obtained while also resolving the thermal noise of the cavity. For reference, see Fig. \ref{fig:dcsquid1}.

\subsection{Back-impedance}

The uncoupled impedance of the hidden photon resonant circuit is
\begin{equation}\label{eq:Zidc}
Z_{i}(\omega)=j\omega L -\frac{j}{\omega C} + R
\end{equation}
As calculated in Ref. \cite{hilbertclarke}, the frequency-dependent back-impedance of the dc SQUID on the transformer circuit is
\begin{equation}\label{eq:dcSQUID_bimt}
\Delta Z_{t} (\omega)=\omega^{2}M_{s}^{2} \left( \frac{1}{j\omega\mathcal{L}_{r}} + \frac{1}{\mathcal{R}_{r}} \right)
\end{equation}
where $\mathcal{L}_{r}$ and $\mathcal{R}_{r}$ are the reduced dynamic input inductance and resistance of the SQUID, respectively, and $M_{s}$ is the mutual inductance between the SQUID inductance and the input coil in the transformer. Thus, the back-impedance of the readout (transformer + dc SQUID) on the hidden photon circuit is
\begin{equation}\label{eq:dcSQUID_bimHP}
\Delta Z_{HP} (\omega)=\frac{\omega^{2}M_{p}^{2}}{j\omega L_{p} + \left(\frac{1}{j \omega L_{sh}} + \frac{1}{j\omega L_{i} + \Delta Z_{t}} \right)^{-1}}
\end{equation}
where $M_{p}$ is the mutual inductance between the hidden photon inductor and the pickup coil (i.e. the sheath). The total impedance of the hidden photon circuit is then, from (\ref{eq:Zidc}),
\begin{equation}\label{eq:Zcdc}
Z_{c}(\omega)=Z_{i}(\omega) + \Delta Z_{HP}(\omega)
\end{equation}
There are two real-valued contributions to the impedance. The first is the intrinsic resistance of the resonant circuit, which arises from quasiparticles and surface dielectric losses, represented by $R$ in equation (\ref{eq:Zidc}). The second is the real part of the back-impedance $\Delta Z_{HP}$, which represents the flow of hidden photon signal energy dissipating in the resistive shunts of the SQUID, manifested in the term $\mathcal{R}_{r}$. At frequencies far below the gap (e.g. the frequencies of interest: 100 Hz-10 MHz) and for temperatures well below the critical temperature, the quasiparticle population will be suppressed; as can be calculated using a Mattis-Bardeen framework. The quality factor is then limited by the back-impedance term (assuming a superconductor with low surface dielectric loss). The loss is fundamentially limited by dissipation in the resistive shunts across the Josephson junctions of the SQUID, $R_s$. In a dc SQUID, the value of the resistive shunts is constrained by the need to operate in a non-hysteretic regime, which requires that \cite{squidhb}
\begin{equation}\label{eq:beta_C}
2 \pi I_0 R_s^2 C/\Phi_0 < 1
\end{equation}
where $I_0$ is the critical current of the junction, $C$ is the junction capacitance, and $\Phi_0$ is the flux quantum. 

The resonance frequency $\omega_{HP}$ is the frequency at which the imaginary part of $Z_{c}(\omega)$ vanishes. The $Q$ can be calculated as follows: 

First, we determine the current through each portion of the circuit as a function of the current $I_{HP}$ in the hidden photon resonator. The current through the pickup coil is
\begin{equation}\label{eq:Ipdc}
I_{p}=\frac{Z_{c}(\omega)-Z_{i}(\omega)}{j\omega M_{p}} I_{HP}
\end{equation}
The current through the inductive shunt is
\begin{equation}\label{eq:Ishdc}
I_{sh}=-\frac{L_{p}I_{p} +  M_{p} I_{HP}}{L_{sh}}
\end{equation}
The current through the input coil
\begin{equation}\label{eq:Iidc}
I_{i}=I_{p}-I_{sh}
\end{equation}
The current through the equivalent-circuit-model dc SQUID dynamic inductance and resistance is
\begin{equation}\label{eq:ILrdc}
I_{\mathcal{L}_{r}}=\frac{\mathcal{R}_{r}}{\mathcal{R}_{r} + j\omega \mathcal{L}_{r}} I_{s}
\end{equation}
\begin{equation}\label{eq:IRrdc}
I_{\mathcal{R}_{r}}=\frac{ j\omega \mathcal{L}_{r}}{\mathcal{R}_{r} + j\omega \mathcal{L}_{r}} I_{s}
\end{equation}
where
\begin{equation}\label{eq:Isdc}
I_{s}=\frac{ L_{sh}I_{sh}- L_{i}I_{i}}{M_{s}}
\end{equation}

We can use the expressions (\ref{eq:Ipdc})-(\ref{eq:Isdc}) to determine the time-averaged stored energy (i.e. averaging fast oscillations to zero) in the inductors and capacitors
\begin{align}\label{eq:ELdc}
E_{L}(\omega) =& \frac{1}{4} \left( L |I_{HP}|^{2} + L_{p}|I_{p}|^{2} + L_{sh}|I_{sh}|^{2} + L_{i} |I_{i}|^{2} + \mathcal{L}_{r} |I_{\mathcal{L}_{r}}|^{2} \right)
\nonumber \\ 
& +\frac{1}{2} \big(M_{p} |I_{HP}| |I_{p}| \cos( \arg (I_{HP} I_{p}^*) ) + M_{s} |I_{i}| |I_{s}| \cos(  \arg (I_{i} I_{s}^*) ) \big)
\end{align}
\begin{equation}\label{eq:ECdc}
E_{C}(\omega)=\frac{1}{4\omega^{2}} \left( \frac{|I_{HP}|^{2}}{C} \right)
\end{equation}
and the power dissipated in the resistors
\begin{equation}\label{eq:PRdc}
P_{diss}(\omega)=\frac{1}{2} \left( |I_{HP}|^{2}R + |I_{\mathcal{R}_{r}}|^{2}\mathcal{R}_{r} \right) 
\end{equation}
The resonant frequency $\omega_{HP}$ of detection circuit is the frequency which maximizes the total energy $E_{L}(\omega)+E_{C}(\omega)$. The quality factor is then
\begin{equation}\label{eq:Qdc}
Q=\frac{\omega_{HP}(E_{L}(\omega_{HP}) + E_{C}(\omega_{HP}))}{P_{diss}(\omega_{HP})}
\end{equation}

The internal Q of a niobium superconducting LC circuit, well below gap frequency and critical temperature, is expected to be larger than one million, so as long as we properly adjust the coupling to the dc SQUID, we should achieve a loaded Q of one million.

\begin{figure}
\vspace{-80pt}
\centerline{\includegraphics[width=12cm]{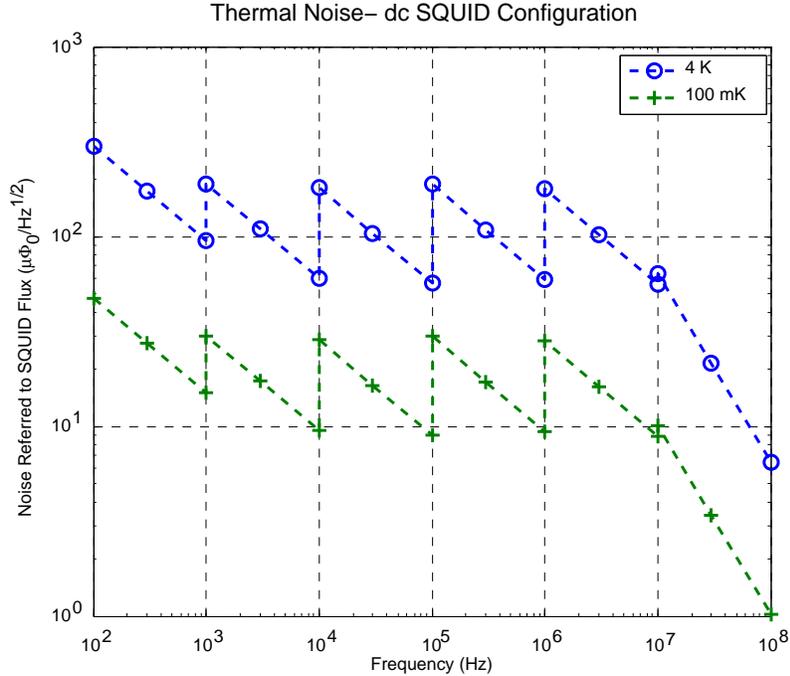}}
\vspace{-80pt}
\caption{ Thermal noise of the cavity, referred to flux noise in the dc SQUID, at 100 mK and 4 K. These numbers were calculated as follows: The input coil had inductance 280 nH, while the pickup coil had inductance 261.5 nH. We set the shunt inductance to a small value (50 pH-5 nH), compared to the input and pickup inductances. For cavity frequencies up to 10 MHz, loss was added to the cavity to bring the internal Q down to approximately 1 million. The shunt inductance was then adjusted to bring the thermal noise at 4 K into the range of $\sim$ 100 $\mu\Phi_{0}/\sqrt{Hz}$. For frequencies up to 10 MHz, this inductance adjustment does not significantly degrade the Q. Above 10 MHz, such an inductance adjustment would significantly degrade the Q, as the coupled Q is closer to one million. At these frequencies, the internal Q was set to 30 million--much higher than the requirement of one million-- and the shunt inductance was adjusted until the coupled Q, and hence the overall Q, was approximately one million. A typical value of shunt inductance for these higher frequencies was 1-3 nH, and the result was a thermal noise in the range of $\sim$ 5-100 $\mu\Phi_{0}/\sqrt{Hz}$ at 4 K. The sawtooth shape of the curve is the result of changing the architecture every decade in frequency, as discussed in the main text; when we change the architecture, we re-optimize the couplings to the SQUID, which results in the jump in flux-referred thermal noise at each decade mark.
} \label{fig:dcapp}
\end{figure}

\subsection{Thermal Noise}

The thermal voltage noise spectral density in the hidden photon resonant circuit is
\begin{equation}
S_{V}(f)=4kTRe(Z_{c})
\end{equation}
where $Z_{c}$ is from (\ref{eq:Zcdc}) and $T$ is the resonator temperature. This translates into a current noise of
\begin{equation}
S_{I}(f)=4kT\frac{Re(Z_{c})}{|Z_{c}|^{2}}
\end{equation}
which induces a current noise in the input coil
\begin{equation}
S_{I,i}(f)=4kT\frac{Re(Z_{c})}{|Z_{c}|^{2}} \frac{M_{p}^{2} L_{sh}^{2}}{((L_{i}+L_{p})L_{sh}+ L_{i}L_{p})^{2}}
\end{equation}
and a thermal flux noise in the SQUID
\begin{equation}
S_{\Phi,Thermal}(f)=4kT\frac{Re(Z_{c})}{|Z_{c}|^{2}} \frac{M_{p}^{2} M_{s}^{2} L_{sh}^{2}}{((L_{i}+L_{p})L_{sh}+ L_{i}L_{p})^{2}}
\end{equation}

To demonstrate that we can simultaneously obtain a loaded quality factor of one million and resolve the cavity thermal noise, we used the above calculations to simulate back-impedance and flux-referred thermal noise in architectures spanning 100 Hz-100 MHz. The transformer couplings and shunt inductor and the internal Q were adjusted to achieve an overall quality factor of one million. The thermal noise as a function of hidden photon frequency is shown in Fig. (\ref{fig:dcapp}) and demonstrates that thermal noise dominates the intrinsic dc SQUID flux noise in the frequency range 100 Hz-10 MHz. The white portion of a typical commercial SQUID flux noise is 3-4 $\mu\Phi_{0}/\sqrt{Hz}$ at 4 K and scales as $\sqrt{T}$ down to approximately 600 mK. \cite{muckdcsquid} As such, below 10 MHz, the thermal noise can be resolved, but above 10 MHz, the intrinsic SQUID flux noise is expected to degrade the sensitivity of our detection circuit.

\section{Microwave ac SQUID circuits for hidden photon readout}
\label{app:acsquidapp}

An overview of our implementation of the microwave SQUID readout technique are presented in \cite{matesphd}. We demonstrate the following in this section:
\begin{itemize}
\item Loaded Qs of one million can be achieved for the hidden photon cavity.
\item Reasonable circuit parameters satisfy the conditions for maximum sensitivity: that the dark photon sidebands lie within the resonant bandwidth of the microwave resonator and the largest detuning from the flux does not move the resonance frequency more than one part in $2Q_{r}$, where $Q_{r}$ is the overall quality factor of the microwave resonator. The overall quality factor is related to the coupled quality factor $Q_{c}$ and internal quality factor $Q_{i}$ by $Q_{r}^{-1}=Q_{c}^{-1}+Q_{i}^{-1}$. The microwave resonators are lithographed from superconducting niobium films and typically possess an internal Q of $10^{6}$. In the geometries and architectures discussed here, the $Q_{c}$ will be less than $10^{2}$. This implies $Q_{r} \approx Q_{c}$, and consequently, we will use the two quality factors interchangeably, referring constraints to the coupled quality factor. 
\item The thermal noise from the hidden photon resonator dominates the feedline amplifier noise and the intrinsic SQUID flux noise.
\end{itemize}

Refer to Fig. \ref{fig:acsquid1} for the definition of important quantities. A few preliminaries first: 

A Josephson junction in an ac SQUID can be modeled as a flux-dependent inductor $L_{J}(\Phi)$, a capacitor $C_{J}$, and resistor $R_{sg}$ (from sub-gap leakage) in parallel, so that its impedance is
\begin{equation}\label{eq:ZSQ0}
Z_{SQ}(\omega)=\left( \frac{1}{j\omega L_{J}(\Phi)} + j\omega C_{J} + \frac{1}{R_{sg}} \right)^{-1}
\end{equation}
where the flux-dependent inductance is 
\begin{equation}\label{eq:LJflux}
L_{J}(\Phi)=L_{J0}\mathrm{sec} \left( 2\pi \frac{\Phi}{\Phi_{0}} \right),\ L_{J0}=\Phi_{0}/2\pi I_{c}
\end{equation}
For the 5 $\mu A$ junction used in present microwave SQUID readouts, $L_{J0}=66$ pH. To prevent magnetic hysteresis, the total loop inductance $L_{1}+L_{2}$ must be less than this value. Typically, $L_{1}=L_{2}=10$ pH, so that $L_{1}+L_{2} \approx \frac{1}{3} L_{J0}$. We operate the microwave resonator at frequencies far below the Josephson plasma frequency $\omega_{J}=1/\sqrt{L_{J0}C_{J}}$ (typically around 50 GHz) and the sub-gap leakage is typically around 1 k$\Omega$. Thus, in this regime, for the purposes of calculating the shift in resonance frequency (but not for the purposes of the calculation of hidden photon resonator Q), the SQUID can be approximated purely as a flux-dependent inductance.

Furthermore, as we will see below the derivative of the resonance frequency as a function of flux varies approximately as $\mathrm{sin} \left( 2\pi \frac{\Phi}{\Phi_{0}} \right)$. For the purposes of the following calculations, we do not bias the dc flux at the maximum of responsivity at $\Phi=\Phi_{0}/4$, as at this flux, the inductive branch of the Josephson junction is open, leading to more current to the resistive branch; this in turn may lead to unacceptable Q degradation in the hidden photon cavity, especially at higher hidden photon frequencies. We instead dc flux bias at a slightly lower responsivity $\Phi=\Phi_{0}/8$. Changes in the SQUID flux come from driving the microwave resonator near the resonance frequency and any external signal (e.g. hidden photon signal). Nevertheless, we note that the actual bias point in an experiment will be determined by a careful experimental characterization of the open-loop performance of the ac SQUID; it may turn out that we may bias closer to maximum responsivity without degrading the Q, which would decrease flux-referred follow-on amplifier noise and yield more flexibility in coupling the hidden photon resonator to the SQUID.

\subsection{Resonant circuit quality factor degradation}
The sub-gap loss $R_{sg}$ in the Josephson junction will degrade the resonant circuit quality factor via the coupling between the SQUID and the  resonant circuit. Here, we determine the quality factor of the coupled and resonant circuit.

For the coupled system, we first determine the impedance of the resonant circuit. Let
\begin{equation}\label{eq:Ziac}
Z_{i}(\omega)=j\omega L - \frac{j}{\omega C} + R
\end{equation}
be the impedance of the uncoupled circuit. The resistance $R$ represents losses in the resonant circuit through surface dielectrics and quasiparticles. We can relate the dark photon source voltage $V_{HP}$, dark photon current $I_{HP}$, and the current through the transformer $I_{t}$ by
\begin{equation}\label{eq:Zc1}
V_{HP}=Z_{i}(\omega)I_{HP} + j\omega M_{p} I_{t}
\end{equation}
where $M_{p}$ is the mutual inductance between the pickup loop and hidden photon resonant circuit.
We can also loop around the inductive transformer and write
\begin{equation}\label{eq:Zc2}
j\omega(L_{i}+L_{p})I_{t} + j\omega M_{s}I_{L_{2}} +j\omega M_{p}I_{HP}=0
\end{equation}
where $M_{s}$ is the mutual inductance between the right-half of the SQUID and input coil.

The voltage across the right- and left-hand sides of the SQUID loop are, respectively,
\begin{equation}\label{eq:Zc4}
V_{L_{1}}=(Z_{SQ}(\omega) + j\omega L_{1})I_{L_{1}}
\end{equation}
\begin{equation}\label{eq:Zc5}
V_{L_{2}}=j\omega M_{s} I_{t} + j\omega L_{2} I_{L_{2}}
\end{equation}
where we approximate the SQUID impedance $Z_{SQ}$ by its value at the dc bias flux $\Phi=\Phi_{0}/8$:
\begin{equation}\label{eq:ZSQ}
Z_{SQ}(\omega)= \left( \frac{1}{\sqrt{2}j\omega L_{J0}} + j\omega C_{J} + \frac{1}{R_{sg}} \right)^{-1}
\end{equation}

The voltages across each of the three legs of the microwave resonator are also equal, so
\begin{equation}\label{eq:Zc6}
I_{R_{r}}R_{r} = \frac{I_{C_{r}}}{j\omega C_{r}} = j\omega L_{r} I_{L_{r}} + (Z_{SQ} + j\omega L_{1})I_{L_{1}}
\end{equation}
Define
\begin{equation}\label{eq:Zc7}
I_{tot}=I_{L_{r}} + I_{R_{r}} + I_{C_{r}}
\end{equation}
as the total current output from the microwave resonator. Then, looping around the microwave resonator circuit,
\begin{equation}\label{eq:Zc8}
j\omega L_{r} I_{L_{r}} + (Z_{SQ} + j\omega L_{1})I_{L_{1}} + \left( Z_{0} + \frac{1}{j\omega C_{c}} \right) I_{tot}=0
\end{equation}
Combining (\ref{eq:Zc6}) and (\ref{eq:Zc7}) yields
\begin{align}
I_{tot} &= I_{L_{r}} + \left( \frac{1}{R_{r}} + j\omega C_{r} \right) \left( j\omega L_{r} I_{L_{r}} + (Z_{SQ}(\omega) + j\omega L_{1}) I_{L_{1}} \right) \nonumber \\
&= \left( 1 + \left( \frac{1}{R_{r}} + j\omega C_{r} \right) \left( Z_{SQ} + j\omega (L_{1} + L_{r}) \right) \right)I_{L_{1}}  + \left( 1 + \left( \frac{1}{R_{r}} + j\omega C_{r} \right) \left( j\omega L_{r} \right) \right)I_{L_{2}}  \label{eq:Zc9}
\end{align}
Plugging this into equation (\ref{eq:Zc8}) gives
\begin{align}
(Z_{SQ}(\omega) &+ j\omega(L_{1} + L_{r})) I_{L_{1}} + \left( Z_{0} + \frac{1}{j\omega C_{c}} \right) \left( 1 + \left( \frac{1}{R_{r}} + j\omega C_{r} \right) \left( Z_{SQ}(\omega) + j\omega (L_{1} + L_{r}) \right) \right)  I_{L_{1}} \nonumber \\
& j\omega L_{r} I_{L_{2}} + \left( Z_{0} + \frac{1}{j\omega C_{c}} \right)\left( 1 + \left( \frac{1}{R_{r}} + j\omega C_{r} \right) \left( j\omega L_{r} \right) \right) I_{L_{2}}=0 \label{eq:Zc10}
\end{align}
which, rearranging, becomes
\begin{equation}\label{eq:Zc11}
I_{L1}=-\frac{ j\omega L_{r} + \left( Z_{0} + \frac{1}{j\omega C_{c}} \right)\left( 1 + \left( \frac{1}{R_{r}} + j\omega C_{r} \right) \left( j\omega L_{r} \right) \right)} { Z_{SQ}(\omega) + j\omega(L_{1} + L_{r}) + \left( Z_{0} + \frac{1}{j\omega C_{c}} \right) \left( 1 + \left( \frac{1}{R_{r}} + j\omega C_{r} \right) \left( Z_{SQ}(\omega) + j\omega (L_{1} + L_{r}) \right) \right) } I_{L_{2}}
\end{equation}
We denote the coefficient in front of $I_{L_{2}}$ by the letter $X$.
Plugging (\ref{eq:Zc11}) into (\ref{eq:Zc4}) and setting it equal to the parallel voltage (\ref{eq:Zc5}) yields
\begin{equation}\label{eq:Zc12}
-j\omega M_{s} I_{t}= I_{L_{2}} ( j\omega L_{2} - X(Z_{SQ}(\omega) + j\omega L_{1}) )
\end{equation}
which gives an expression for $I_{L_{2}}$ that can be substituted into (\ref{eq:Zc2}):
\begin{equation}\label{eq:Zc13}
j\omega(L_{i}+L_{p})I_{t} + \frac{ \omega^{2} M_{s}^{2} }{ j\omega L_{2} - X(Z_{SQ}(\omega) + j\omega L_{1})} I_{t} +j\omega M_{p}I_{HP}=0
\end{equation}
which in turn gives an expression for $I_{t}$ that can be substituted into (\ref{eq:Zc1}):
\begin{equation}\label{eq:Zc14}
V_{HP}=Z_{i}(\omega)I_{HP} + \frac{\omega^{2}M_{p}^{2}}{ j\omega(L_{i}+L_{p}) + \frac{ \omega^{2} M_{s}^{2} }{ j\omega L_{2} - X(Z_{SQ}(\omega) + j\omega L_{1})} }I_{HP}
\end{equation}
so the coupled impedance is
\begin{equation}\label{eq:Zcac}
Z_{c}(\omega)=Z_{i}(\omega) + \frac{\omega^{2}M_{p}^{2}}{ j\omega(L_{i}+L_{p}) + \frac{ \omega^{2} M_{s}^{2} }{ j\omega L_{2} - X(Z_{SQ}(\omega) + j\omega L_{1})} }
\end{equation}

The coupled impedance will contain three resistive contributions: one from quasiparticle/dielectric losses in the superconducting LC circuit, one from coupling to the sub-gap loss in the SQUID, and one from coupling to the $Z_{0}=50 \ \Omega$ feedline termination resistor. Because of the very high impedance of the capacitive coupling to the feedline, the third contribution is subdominant to the first two. We operate at temperatures far below $T_{c}$ and in a regime where the microwave SQUID circuit is strongly coupled to the resonant circuit, which allows internal Qs of much higher than one million. This is confirmed with a straightforward Mattis-Bardeen computation.  As such, we are likely dominated by couplings to the sub-gap loss in the SQUID. 

We calculate the resonance frequency and quality factor in a manner similar to the dc SQUID case. The resonance frequency $\omega_{HP}$ is the frequency at which the imaginary component of $Z_{c}$ vanishes. To calculate Q, we work backwards through (\ref{eq:Ziac})-(\ref{eq:Zcac}) and evaluate the current in each portion of the detection circuit in terms of the dark photon induced current $I_{HP}$.

The current in the transformer coil is
\begin{equation}\label{eq:Itac}
I_{t}=\frac{Z_{c}(\omega)-Z_{i}(\omega)}{j\omega M_{p}}I_{HP}
\end{equation}
The current in the left-hand side of the SQUID loop is
\begin{equation}\label{eq:IL2ac}
I_{L_{2}}=-\frac{j\omega M_{s}}{j\omega L_{2} - X(Z_{SQ}+j\omega L_{1})}I_{t}
\end{equation}
The current in the right-hand side of the SQUID loop is
\begin{equation}\label{eq:IL1ac}
I_{L_{1}}=XI_{L_{2}}
\end{equation}
These equations, in turn, yield the current through the microwave resonator inductor $L_{r}$
\begin{equation}\label{eq:ILrac}
I_{L_{r}}=I_{L_{1}}+I_{L_{2}}
\end{equation}
the current through the microwave resonator capacitor $C_{r}$
\begin{equation}\label{eq:ICrac}
I_{C_{r}}=j\omega C_{r} (j\omega L_{r} I_{L_{r}} + (Z_{SQ} + j\omega L_{1}) I_{L_{1}})
\end{equation}
and the current through the resistor $R_{r}$
\begin{equation}\label{eq:IRrac}
I_{R_{r}}=\frac{I_{C_{r}}}{j\omega R_{r}C_{r}}
\end{equation}
Equation (\ref{eq:IL1ac}) also gives the approximate current in the inductive, capacitive, and resistive branches of the Josephson junction.
\begin{equation}\label{eq:ILJac}
I_{L_{J}}=\frac{Z_{SQ}}{\sqrt{2}j\omega L_{J0}}I_{L_{1}}
\end{equation}
\begin{equation}\label{eq:ICJac}
I_{C_{J}}=j\omega C_{J}Z_{SQ}I_{L_{1}}
\end{equation}
\begin{equation}\label{eq:IRJac}
I_{R_{sg}}=\frac{Z_{SQ}}{R_{sg}}I_{L_{1}}
\end{equation}
Finally, the total current output from the microwave resonator is that from (\ref{eq:Zc7}).

Now, combining equations (\ref{eq:Zc7}), (\ref{eq:Itac})-(\ref{eq:IRJac}), the total, time-averaged inductive and capacitive energies are (using the zeroth-order Josephson energy as an approximation)
\begin{align}\label{eq:ELac}
E_{L}(\omega)=& \frac{1}{4}(L|I_{HP}|^{2} + (L_{i}+L_{p})|I_{t}|^{2} + L_{J0}|I_{L_{J}}|^{2} + L_{1} |I_{L_{1}}|^{2} + L_{2} |I_{L_{2}}|^{2} + L_{r}|I_{L_{r}}|^{2})
\nonumber \\
& + \frac{1}{2} \big(M_{p} |I_{HP}| |I_{t}| \cos( \arg (I_{HP} I_{t}^*) ) + M_{s} |I_{t}| |I_{L_{2}}| \cos( \arg (I_{t} I_{L_2}^*) ) \big)
\end{align}
\begin{equation}\label{eq:ECac}
E_{C}(\omega)= \frac{1}{4\omega^{2}} \left( \frac{|I_{HP}|^{2}}{C} + \frac{|I_{C_{J}}|^{2}}{C_{J}} + \frac{|I_{C_{r}}|^{2}}{C_{r}} + \frac{|I_{tot}|^{2}}{C_{c}} \right)
\end{equation}
and the power dissipated in the hidden photon detection circuit is
\begin{equation}
P_{diss}(\omega)= \frac{1}{2}(|I_{HP}|^{2}R+|I_{R_{sg}}|^{2}R_{sg} + |I_{R_{r}}|^{2}R_{r}+ |I_{tot}|^{2}Z_{0})
\end{equation}

The resonant frequency $\omega_{HP}$ of detection circuit is the frequency which maximizes the total energy $E_{L}(\omega)+E_{C}(\omega)$. The quality factor is then
\begin{equation}\label{eq:Qac}
Q=\frac{\omega_{HP}(E_{L}(\omega_{HP}) + E_{C}(\omega_{HP}))}{P_{diss}(\omega_{HP})}
\end{equation}

As in the dc SQUID case, the internal Q of the hidden photon LC resonator is expected to be greater than one million. Therefore, by using the proper couplings, an overall Q of one million can be obtained.

\begin{figure}
\vspace{-20pt}
\centerline{\includegraphics[width=22cm]{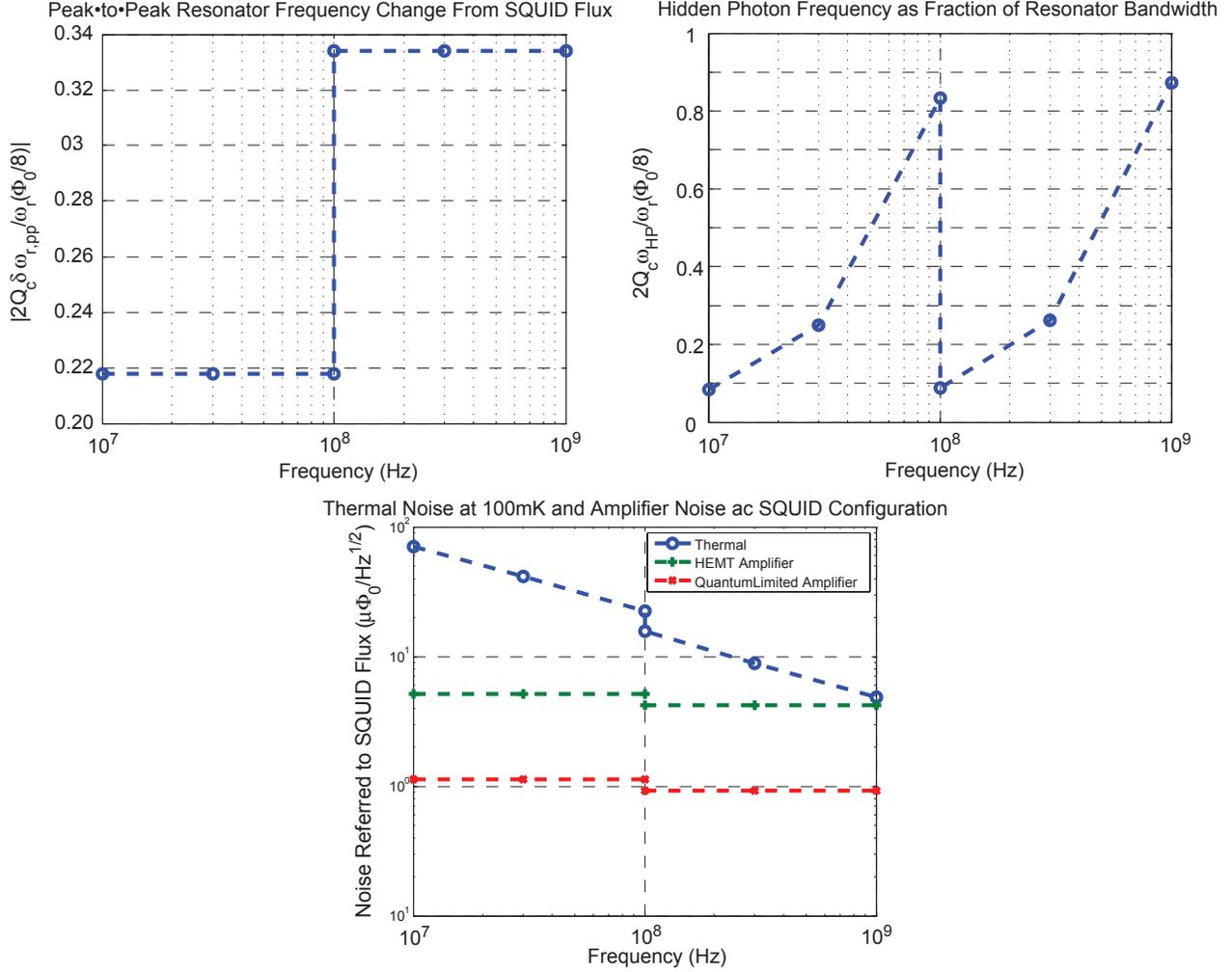}}
\vspace{-40pt}
\caption{ Top left: Maximum resonant frequency change from SQUID flux, as a fraction of microwave resonator bandwidth. Top right: Hidden photon resonant frequency, as a fraction of microwave resonator bandwidth. Bottom: Flux-referred thermal noise of the cavity at 100 mK plotted with the flux-referred HEMT noise and quantum-limited parametric amplifier noise over 10 MHz- 1 GHz. The numbers were calculated as follows: The input coil and pickup coil inductances were the same as for the dc SQUID. Two architectures were used here, one for 10-100 MHz and the other for 100 MHz-1 GHz. First, the microwave resonator frequency and coupled Q were chosen so that the hidden photon resonant frequency would be a fraction of the microwave resonator bandwidth. The resonator frequencies were 8.3 and 23.6 GHz for the lower-frequency and higher-frequency architectures, respectively, with coupled Q of 32 and 10. The linear increase in hidden photon frequency with constant resonator frequency explains the exponential-like curves on the semi-log plot in the top left panel. Second, the microwave resonator inductance $L_{r}$ was chosen so that the peak-to-peak frequency modulation from the SQUID flux, i.e. the quantity $|\delta \omega_{r,pp}|$, would be at least 1/5 of the resonator bandwidth. This sets a moderate level of SQUID-resonator coupling and readout sensitivity. For realistic fabrication parameters, we must decrease $L_{r}$ to obtain a higher microwave resonator frequency, which results in the increase in peak-to-peak modulation fraction for the higher frequency architecture. To achieve a loaded Q of one million, up to 100 MHz, the coupling between the pickup coil and the hidden photon resonator inductance was set at $k_{p}=0.5$, and the coupling between the SQUID loop and the input coil was set at $k_{s}=0.6$ (typical coupling for a microwave SQUID). The internal loss was adjusted until a Q of one million was achieved. Above 100 MHz, the coupled Q was lower than one million for these values of $k_{s}$ and $k_{p}$ ; thus, $k_{p}$ was reduced to 0.35. This results in the reduction in flux-referred thermal noise at 100 MHz when we change architectures. The resonator was driven at the highest possible power, corresponding to $\gamma_{c}=1$, to ensure the best possible flux-referred amplifier noise. The higher peak-to-peak flux modulation for the higher frequency architecture means greater SQUID-to-amplifier coupling and slightly reduced flux-referred amplifier noise.
} \label{fig:acapp}
\end{figure}

\subsection{Resonance Frequency and Coupled Quality Factor of Microwave SQUID Resonator}
Here, we calculate the resonance frequency and coupled-$Q$ of the microwave SQUID resonator, as a function of flux through the SQUID. The impedance of the resonator is
\begin{equation}\label{eq:ZR1}
Z_{R}(\omega)=\frac{1}{j\omega C_{c}} + \left( \frac{1}{j\omega L_{r} + \frac{(Z_{SQ}(\omega) + j\omega L_{1})(j\omega L_{2})}{j\omega (L_{1} + L_{2}) + Z_{SQ}(\omega)}} + j\omega C_{r} + \frac{1}{R_{r}} \right)^{-1}
\end{equation}
We operate the resonator at frequencies far below the Josephson plasma frequency (typically around 50 GHz); in this regime, the SQUID can be modeled purely as the flux-dependent inductance of equation (\ref{eq:LJflux}). Thus, the total inductance of the microwave resonator is
\begin{equation}\label{eq:Lt}
L_{t}(\Phi)=L_{r} + \frac{L_{2}(L_{1}+L_{J}(\Phi))}{L_{1}+L_{2}+L_{J}(\Phi)}
\end{equation}
The first term is the native inductance of the microwave resonator, and the second term combines the SQUID loop and Josephson inductances. (\ref{eq:ZR1}) simplifies to
\begin{align}
Z_{R}(\omega,\Phi) &= \frac{1}{j\omega C_{c}} + \left( \frac{1}{j\omega L_{t}(\Phi)} + j\omega C_{r} + \frac{1}{R_{r}} \right)^{-1} \nonumber \\
&= \frac{1}{j\omega C_{c}} + \frac{j\omega L_{t}(\Phi)}{\frac{j\omega L_{t}(\Phi)}{R_{r}} + (1-\omega^{2}L_{t}(\Phi)C_{r})} \nonumber \\
&= \frac{1}{j\omega C_{c}} + \frac{j\omega L_{t}(\Phi) \left(  1-\omega^{2}L_{t}(\Phi)C_{r} - \frac{j\omega L_{t}(\Phi)}{R_{r}} \right) }{ (1-\omega^{2}L_{t}(\Phi)C_{r})^{2} + \frac{\omega^{2} L_{t}^{2}(\Phi)}{R_{r}^{2}} } \nonumber \\
&= \frac{j\omega L_{t}(\Phi) \left(  1-\omega^{2}L_{t}(\Phi)C_{r} - \frac{j\omega L_{t}(\Phi)}{R_{r}} \right) + \frac{1}{j\omega C_{c}} \left( (1-\omega^{2}L_{t}(\Phi)C_{r})^{2} + \frac{\omega^{2} L_{t}^{2}(\Phi)}{R_{r}^{2}}  \right) }{ (1-\omega^{2}L_{t}(\Phi)C_{r})^{2} + \frac{\omega^{2} L_{t}^{2}(\Phi)}{R_{r}^{2}} } \label{eq:ZR2}
\end{align}
At the resonance frequency, the imaginary part of the impedance vanishes, so we calculate the flux-dependent resonance frequency $\omega_{r}(\Phi)$ to be
\begin{equation}\label{eq:omegaresflux1}
\omega_{r}(\Phi)=\frac{1}{\sqrt{L_{t}(\Phi)(C_{r}+C_{c})}}
\end{equation}

Though the microwave resonator may have high internal $Q$, the power dissipation is dominated by loss on the feedline, similar to how the hidden photon resonant circuit $Q$ is dominated by the sub-gap loss in the SQUID; the coupled quality factor $Q_{c}$ accounts for the loss from the feedline coupling. We calculate the coupled $Q$ by considering the reflection coefficient for a traveling wave on the feed line with applied flux $\Phi$ in the SQUID:
\begin{equation}\label{eq:gamma1}
\Gamma = \frac{Z_{R}(\omega,\Phi) - Z_{0}}{Z_{R}(\omega,\Phi) + Z_{0}} = 1 - \frac{2}{\frac{Z_{R}(\omega,\Phi)}{Z_{0}} + 1}
\end{equation}

Off resonance, the imaginary part of $Z_{R}$ will dominate the real part, assuming the resonator circuit is low-loss. Thus, we can parameterize
\begin{equation}\label{eq:tantheta}
\frac{Z_{R}(\omega,\Phi)}{Z_{0}} = j\mathrm{tan}(\theta)
\end{equation}
so that
\begin{equation}\label{eq:gamma2}
\Gamma=1 - \frac{2}{1 + j\mathrm{tan}(\theta)} = 1 - \frac{2(1 - j\mathrm{tan}(\theta))}{1 + \mathrm{tan}^{2}(\theta)} = -\mathrm{cos}(2\theta) + j\mathrm{sin}(2\theta)
\end{equation}
The reflection coefficient thus traces out a circle in the complex plane centered at the origin. Resonance occurs when $\Gamma \approx -1$, at which the phase of the reflection coefficient is $-\pi$, and on the circle, $\theta=-\pi$. The frequencies determining the FWHM correspond to a phase (of $\Gamma$) of $\pm \pi/2$. We write equations determining these frequencies.

From (\ref{eq:ZR2}), in the limit of high internal $Q$, and for frequencies near $\omega_{r}(\Phi)$,
\begin{equation}\label{eq:ImZR}
Im(Z_{R}(\omega,\Phi))=\frac{1}{\omega C_{c} \left(1- \omega^{2}L_{t}(\Phi)C_{r} \right) }\left( \omega^{2}L_{t}(\Phi)C_{c} - (1-\omega^{2}L_{t}(\Phi)C_{r} ) \right)
\end{equation}
In the case of $-\pi/2$, $Im(Z_{R}(\omega,\Phi))=-Z_{0}$. Let $\omega_{1}$ be the frequency that solves this equation. Setting (\ref{eq:ImZR}) equal to $-Z_{0}$ yields
\begin{equation}\label{eq:ImZRomega1}
1-\omega_{1}^{2}L_{t}(\Phi)(C_{r}+C_{c}) = \omega_{1}C_{c}Z_{0}(1-\omega_{1}^{2}L_{t}(\Phi)C_{r})
\end{equation}
In the case of $+\pi/2$, $Im(Z_{R}(\omega,\Phi))=+Z_{0}$. Let $\omega_{2}$ be the frequency that solves this equation. Then $\omega_{2}$ satisfies
\begin{equation}\label{eq:ImZRomega2}
1-\omega_{2}^{2}L_{t}(\Phi)(C_{r}+C_{c}) = \omega_{2}C_{c}Z_{0}(1-\omega_{2}^{2}L_{t}(\Phi)C_{r})
\end{equation}
Typically, equations (\ref{eq:ImZRomega1}) and (\ref{eq:ImZRomega2}) would be solved numerically, and the coupled Q would be determined by $Q_{c}=\omega_{r}(\Phi)/(\omega_{2}-\omega_{1})$. However, in readout architectures for hidden photons below 100 MHz, we typically have $\omega_{r}(\Phi)C_{c}Z_{0} \ll 1$ and $|\omega_{1}-\omega_{r}(\Phi)|, |\omega_{1}-\omega_{r}(\Phi)| \ll \omega_{r}(\Phi)$. As such, the equations can be solved by linearizing.
Plugging $\omega_{1}=\omega_{r}(\Phi)-\delta\omega_{1}$ into (\ref{eq:ImZRomega1}) and linearizing yields
\begin{equation}\label{eq:ImZRdomega1}
\delta\omega_{1}=\frac{1}{2}\omega_{r}(\Phi)C_{c}Z_{0}\omega_{r}(\Phi)\frac{C_{c}}{C_{r}+C_{c}} 
\end{equation}
Plugging $\omega_{2}=\omega_{r}(0)+\delta\omega_{2}$ into (\ref{eq:ImZRomega2}) and linearizing yields
\begin{equation}\label{eq:ImZRdomega2}
\delta\omega_{2}=\frac{1}{2}\omega_{r}(\Phi)C_{c}Z_{0}\omega_{r}(\Phi)\frac{C_{c}}{C_{r}+C_{c}}\end{equation}
The bandwidth is then
\begin{equation}\label{eq:BW}
BW=\omega_{2}-\omega_{1} = \delta\omega_{1} + \delta\omega_{2} = \omega_{r}(\Phi)C_{c}Z_{0}\omega_{r}(\Phi)\frac{C_{c}}{C_{r}+C_{c}}
\end{equation}
which yields a coupled-$Q$ of
\begin{equation}\label{eq:Qc1}
Q_{c}=\frac{\omega_{r}(\Phi)}{BW}=\frac{C_{r}+C_{c}}{\omega_{r}(\Phi)C_{c}^{2}Z_{0}}=\frac{1}{\omega_{r}(\Phi)C_{c}Z_{0}\omega_{r}(\Phi)^{2}L_{t}(\Phi)C_{c}}
\end{equation}
Note that the coupled-Q is dependent on the flux, $Q_{c}=Q_{c}(\Phi)$. However, in all architectures presented here $Q_{c}(\Phi_{0}/8)$ will be at least $\sim 10$. As discussed in the next section, we will constrain the peak-to-peak modulation of the resonance frequency to be less than one part in $2Q_{c}(\Phi_{0}/8)$. Thus, the coupled Q will vary with flux by less than one part in 20, and we can take the coupled Q to be a constant, $Q_{c}=Q_{c}(\Phi=\Phi_{0}/8)$.

\subsection{Resonant Frequency Modulation by SQUID Flux}

As discussed in the previous section, the resonance frequency is dependent upon the flux through the SQUID. The flux from the dark photon signal and the microwave resonator drive will change the resonance frequency. To ensure maximum readout sensitivity, we require that the microwave resonator drive frequency be within the bandwidth of the resonance for all possible SQUID flux. This requirement can be satisfied by constraining the peak-to-peak modulation of the resonance frequency to vary by less than one part in $2Q_{c}=2Q_{c}(\Phi_{0}/8)$. The total inductance of the microwave resonator, given by equation (\ref{eq:Lt}), is dominated by the native inductance $L_{r}$ of the resonator, so we can expand the resonance frequency (\ref{eq:omegaresflux1}) as
\begin{equation}\label{eq:omegaresfluxTE}
\omega_{r}(\Phi) \approx \frac{1}{\sqrt{L_{r}(C_{r}+C_{c})}} \left( 1 - \frac{1}{2} \frac{L_{2}(L_{1}+L_{J}(\Phi))}{L_{r} (L_{1}+L_{2}+L_{J}(\Phi))} \right)
\end{equation}
The maximum resonance frequency occurs at integer multiples of the flux quantum and the minimum resonance frequency occurs at half-integer multiples of the flux quantum. Thus, the peak-to-peak modulation is
\begin{align}
\delta\omega_{r,pp} &= \omega_{r}(0) - \omega_{r}(\Phi_{0}/2)=  -\frac{1}{2}\frac{L_{2}}{L_{r}} \frac{1}{\sqrt{L_{r}(C_{r}+C_{c})}} \left( \frac{L_{1} + L_{J0}}{L_{1}+L_{2}+L_{J0}}  - \frac{L_{1} - L_{J0}}{L_{1}+L_{2}-L_{J0}}  \right ) \nonumber \\
&= -\frac{1}{2}\frac{L_{2}}{L_{r}} \frac{1}{\sqrt{L_{r}(C_{r}+C_{c})}} \left( \frac{ (L_{1} + L_{J0})(L_{1}+L_{2}-L_{J0}) - (L_{1} - L_{J0}) (L_{1}+L_{2}+L_{J0}) }{ (L_{1}+L_{2})^{2} - L_{J0}^{2} } \right) \nonumber \\
&= \frac{1}{\sqrt{L_{r}(C_{r}+C_{c})}} \frac{ L_{2}^{2}L_{J0} }{ L_{r} (L_{J0}^{2} - (L_{1}+L_{2})^{2} ) } \approx \omega_{r}(\Phi_{0}/8) \frac{ L_{2}^{2}L_{J0} }{ L_{r} (L_{J0}^{2} - (L_{1}+L_{2})^{2} ) }\label{eq:domegafluxmax}
\end{align}
Thus, we require
\begin{equation}\label{eq:fluxmodreq}
\frac{\delta \omega_{r,pp}}{\omega_{r}(\Phi_{0}/8)}=\frac{ L_{2}^{2}L_{J0} }{ L_{r} (L_{J0}^{2} - (L_{1}+L_{2})^{2} ) } < \frac{1}{2Q_{c}}
\end{equation}
where the coupled-Q $Q_{c}$ is given in (\ref{eq:Qc1}) by the value at flux $\Phi_{0}/8$. 

As discussed in the main text, a Fourier transform of the phase of the reflected signal will have sidebands separated from the drive frequency; for a positive detection of dark photons, the distance of separation is the dark photon frequency. To maximize sensitivity, these sidebands must be within the bandwidth of the microwave resonator:
\begin{equation}\label{eq:HPSBreq}
\frac{\omega_{HP}}{\omega_{r}(\Phi_{0}/8)} < \frac{1}{2Q_{c}}
\end{equation}
The top two panels of Fig. \ref{fig:acapp} show that equations (\ref{eq:fluxmodreq}) and (\ref{eq:HPSBreq}) are readily satisfied for architectures in the 100 MHz-1 GHz range.

Also, note that from (\ref{eq:Lt}) and (\ref{eq:omegaresflux1}), we can calculate the flux-to-resonance frequency responsivity.
\begin{align}
\frac{d\omega_{r}}{d\Phi} =& -\frac{1}{2}\frac{L_{2}}{L_{r}} \frac{1}{\sqrt{L_{r}(C_{r}+C_{c})}} \frac{d}{d\Phi} \left( 1- \frac{L_{2}\mathrm{cos} \left( 2\pi \frac{\Phi}{\Phi_{0}} \right)}{(L_{1}+L_{2})\mathrm{cos} \left( 2\pi \frac{\Phi}{\Phi_{0}} \right) + L_{J0}} \right) \nonumber \\
= &\frac{1}{2}\frac{L_{2}}{L_{r}} \frac{1}{\sqrt{L_{r}(C_{r}+C_{c})}} \frac{2\pi}{\Phi_{0}}\nonumber \\
& \frac{ \left( -L_{2} \mathrm{sin} \left( 2\pi \frac{\Phi}{\Phi_{0}} \right) \right) \left( (L_{1}+L_{2})\mathrm{cos} \left( 2\pi \frac{\Phi}{\Phi_{0}} \right) + L_{J0} \right) - \left( L_{2} \mathrm{cos} \left( 2\pi \frac{\Phi}{\Phi_{0}} \right) \right) \left( -(L_{1}+L_{2})\mathrm{sin} \left( 2\pi \frac{\Phi}{\Phi_{0}} \right) \right)}{ \left( (L_{1}+L_{2})\mathrm{cos} \left( 2\pi \frac{\Phi}{\Phi_{0}} \right) + L_{J0} \right)^{2}} \nonumber \\
=& \frac{1}{2}\frac{L_{2}}{L_{r}} \frac{1}{\sqrt{L_{r}(C_{r}+C_{c})}} \frac{2\pi}{\Phi_{0}} \frac{-L_{2}L_{J0}\mathrm{sin} \left( 2\pi \frac{\Phi}{\Phi_{0}} \right)}{ \left( (L_{1}+L_{2})\mathrm{cos} \left( 2\pi \frac{\Phi}{\Phi_{0}} \right) + L_{J0} \right)^{2}} \label{eq:domegardPhi}
\end{align}
As mentioned above, $L_{1}+L_{2} \approx \frac{1}{3} L_{J0}$, so the responsivity is approximately a sine function with period $\Phi_{0}$.

\subsection{Constraints On Feedline Power}
There are two significant constraints on feedline power. 

First, the current through the junction should not exceed $I_{c}$.
The current transmission coefficient is
\begin{equation}\label{eq:TI}
T_{I}=1-\Gamma= \frac{2Z_{0}}{Z_{0} + Z_{R}(\omega, \Phi)}
\end{equation}
where $\Gamma$ is defined in equation (\ref{eq:gamma1}). Let $I_{in}$ be the feedline current incident upon the resonator. Then, the current through the inductive part of the resonator is, assuming we are driving near resonance and the resonator has high internal Q,
\begin{equation}\label{eq:FPIL}
I_{L}=\frac{T_{I} I_{in}}{1-\omega^{2}L_{t}(\Phi)C_{r}}
\end{equation}
which, after evaluating at a drive frequency $\omega_{r}(\Phi_{0}/8)$, gives a current through the junction of
\begin{equation}\label{eq:FPIJ}
I_{J}(\Phi)=\frac{L_{2}}{L_{1}+L_{2}+L_{J}(\Phi)}\frac{T_{I} I_{in}}{1-\omega_{r}(\Phi_{0}/8)^{2}L_{t}(\Phi)C_{r}}
\end{equation}
We can express our drive strength as a dimensionless quantity. Using (\ref{eq:FPIJ}), along with $T_{I}\approx 2$ near resonance, and evaluating at $\Phi=\Phi_{0}/8$, we define
\begin{equation}\label{eq:drivestr}
\alpha=\frac{|I_{J}(\Phi_{0}/8)|}{I_{c}}=\frac{2}{1-\omega^{2}L_{t}(\Phi_{0}/8)C_{r}}\frac{L_{2}}{L_{1}+L_{2}+L_{J0}\sqrt{2}} \frac{|I_{in}|}{I_{c}}
\end{equation}
and require that it be less than some value $\alpha_{c}$. This gives a maximum rms feedline power of
\begin{equation} \label{eq:FPlim2}
P_{max,I_{c}} = \frac{1}{2} |I_{in}|^{2} Z_{0} = \left( \alpha_{c} I_{c} \frac{L_{1}+L_{2}+L_{J0}\sqrt{2}}{L_{2}}  (1-\omega_{r}(\Phi_{0}/8)^{2}L_{t}(\Phi_{0}/8)C_{r})  \right)^{2}   \frac{Z_{0}}{8}
\end{equation}

Second, we require that the microwave drive not change the flux (from the dc value of $\Phi_{0}/8$) so as to degrade the flux-to-resonance frequency responsivity. The responsivity goes approximately as a sine function, per our discussion in the previous section. A flux of $\Phi_{0}/8$ corresponds to a phase of $\pi/4$. $\mathrm{sin}(\pi/4)=\sqrt{2}/2$ and the responsivity drops by a factor of 2 (to $\sqrt{2}/4$) at approximately $\pi/8$. Thus, the input current can only change the phase by at most $\pi/8$, or the flux by $\Phi_{0}/16$. The flux through the SQUID from the microwave drive has a maximum absolute value of approximately
\begin{equation}\label{eq:phimax}
|\Phi_{max}| \approx |L_{2}(I_{L}-I_{J}(\Phi_{0}/8)) - L_{1}I_{J}(\Phi_{0}/8)|= \Bigg| \frac{L_{2}L_{J0}\sqrt{2}}{L_{1}+L_{2}+L_{J0}\sqrt{2}} \frac{T_{I} I_{in}}{1-\omega_{r}(\Phi_{0}/8)^{2}L_{t}(\Phi)C_{r}} \Bigg|\end{equation}
We define this quantity as a fraction of $\Phi_{0}/16$, similar to the first case. Using $T_{I} \approx 2$, 
\begin{equation}\label{eq:drivestr2}
\gamma=\frac{16\Phi_{max}}{\Phi_{0}}=\frac{16\Phi_{max}}{2 \pi L_{J0} I_{c}}=  \frac{8\sqrt{2}}{\pi} \frac{2}{1-\omega_{r}(\Phi_{0}/8)^{2}L_{t}(\Phi_{0}/8)C_{r}}\frac{L_{2}}{L_{1}+L_{2}+L_{J0}\sqrt{2}} \frac{|I_{in}|}{I_{c}}
\end{equation}
We require that this be less than some value $\gamma_{c}$, which gives a maximum rms feedline power of\begin{equation} \label{eq:FPlim3}
P_{max,\Phi} = \frac{1}{2} |I_{in}|^{2} Z_{0}= \left( \frac{\pi}{8\sqrt{2}} \gamma_{c} I_{c} \frac{L_{1}+L_{2}+L_{J0}\sqrt{2}}{L_{2}}  (1-\omega_{r}(\Phi_{0}/8)^{2}L_{t}(\Phi_{0}/8)C_{r})  \right)^{2}   \frac{Z_{0}}{8}
\end{equation}
For equal values of $\alpha_{c}$ and $\gamma_{c}$, the second constraint is more stringent. As such, the second constraint is the one that we will utilize. 
Note resonator bifurcation as a constraint on feedline power is not considered here. This is because the dominant inductance in the microwave SQUID circuit, $L_{r}$, is linear, and therefore, though the resonator will show some nonlinearity, it will not bifurcate.

\subsection{Amplifier and Thermal Noise}
The voltage incident on the feedline amplifier is the amplitude of the voltage wave that is reflected back from the resonator
\begin{equation}\label{eq:VR}
V_{R}=\Gamma V_{in}
\end{equation}
We calculate the voltage-to-flux responsivity $dV_{R}/d\Phi$ at $\Phi_{0}/8$, which determines the amplifier noise. Expanding the resonator impedance near $\omega_{r}(\Phi_{0}/8)$, we find
\begin{equation}\label{eq:ZRTE}
Z_{R}(\omega) \approx \frac{\partial Z_{R}}{\partial \omega} (\omega-\omega_{r}(\Phi_{0}/8))
\end{equation}
Plugging into the above equation for $\omega_{1}$ and $\omega_{2}$, as defined in Section IIIA, we find
\begin{equation}\label{eq:dZRQc1}
2iZ_{0}\approx \frac{\partial Z_{R}}{\partial \omega}(\omega_{2}-\omega_{1}) \approx \frac{\partial Z_{R}}{\partial \omega} \frac{\omega_{r}(\Phi_{0}/8)}{Q_{c}}
\end{equation}
so
\begin{equation}\label{eq:dZRQc2}
\frac{\partial Z_{R}}{\partial \omega} = \frac{2iQ_{c}Z_{0}}{\omega_{r}(\Phi_{0}/8)}
\end{equation}
For small shifts, shifting the drive tone away from the resonance is the reverse of shifting the resonance away from the drive tone, so, using (\ref{eq:TI}) and (\ref{eq:dZRQc2}),
\begin{equation}\label{eq:dTIdomegar}
\frac{d}{d\omega_{r}} \left( \frac{1}{T_{I}} \right) \approx -\frac{d}{d\omega} \left( \frac{1}{T_{I}} \right) = -\frac{1}{2Z_{0}}\frac{dZ_{R}}{d\omega} = -\frac{iQ_{c}}{\omega_{r}(\Phi/8)} = -\frac{1}{T_{I}^{2}}\frac{dT_{I}}{d\omega_{r}} = +\frac{1}{T_{I}^{2}}\frac{d\Gamma}{d\omega_{r}}
\end{equation}
With $T_{I}\approx 2$, we find
\begin{equation}\label{eq:dGdomegar}
\frac{d\Gamma}{d\omega_{r}}=-\frac{4iQ_{c}}{\omega_{r}(\Phi_{0}/8)}
\end{equation}
Evaluating equation (\ref{eq:domegardPhi}) at $\Phi_{0}/8$,
\begin{equation}\label{eq:domegardPhi2}
\frac{d\omega_{r}}{d\Phi}  =  - \frac{1}{2}\frac{L_{2}}{L_{r}} \frac{1}{\sqrt{L_{r}(C_{r}+C_{c})}} \frac{2\pi}{\Phi_{0}} \frac{L_{2}L_{J0}\frac{1}{\sqrt{2}}}{ \left( (L_{1}+L_{2}) \frac{1}{\sqrt{2}} + L_{J0} \right)^{2}}
\end{equation}
Combining (\ref{eq:dGdomegar}) and (\ref{eq:domegardPhi2}), we find
\begin{equation}\label{eq:dVRdPhi}
\frac{dV_{R}}{d\Phi}=V_{in}\frac{d\Gamma}{d\Phi}= V_{in} \frac{2iQ_{c}}{\omega_{r}(\Phi_{0}/8)}\frac{L_{2}}{L_{r}} \frac{1}{\sqrt{L_{r}(C_{r}+C_{c})}} \frac{2\pi}{\Phi_{0}}  \frac{L_{2}L_{J0}\frac{1}{\sqrt{2}}}{ \left( (L_{1}+L_{2}) \frac{1}{\sqrt{2}} + L_{J0} \right)^{2}}
\end{equation}
$\omega_{r}(\Phi_{0}/8) \approx \frac{1}{\sqrt{L_{r}(C_{r}+C_{c})}}$, so
\begin{equation}\label{eq:dVRdPhi2}
\frac{dV_{R}}{d\Phi}=\frac{2\pi}{\Phi_{0}}  2iQ_{c} V_{in} \frac{L_{2}}{L_{r}} \frac{L_{2}L_{J0}\frac{1}{\sqrt{2}}}{ \left( (L_{1}+L_{2}) \frac{1}{\sqrt{2}} + L_{J0} \right)^{2}}
\end{equation}

The amplifier noise referred to the SQUID flux is then
\begin{equation}\label{eq:SPhiHEMT}
S_{\Phi,Amp}=S_{V}\left| \frac{dV_{R}}{d\Phi} \right|^{-2}=4kT_{N}Z_{0}\left| \frac{dV_{R}}{d\Phi} \right|^{-2}
\end{equation}
For a HEMT, the noise temperature is typically $T_{N}=1$ K/GHz, while for a quantum-limited parametric amplifier, adding 1/2 photons of noise, the noise temperature is $T_{N}=0.048$ K/GHz. Note that the HEMT noise spectral density referred to SQUID flux goes down as $V_{in}^{-2}$, which is constrained by the feedline power considerations from the previous section.

Another source of noise in this readout scheme is the intrinsic flux noise of the dissipationless rf SQUID. In the regime of interest (above 10 MHz), this noise is expected to be subdominant to the microwave amplifier noise. Also, note that we evade the low-frequency two-level system noise that has been observed universally in superconducting microresonators. \cite{matesphd}

The thermal noise of the resonant circuit referred to the SQUID is the $L_{sh} \rightarrow \infty$ limit of the dc SQUID case:
\begin{equation}\label{eq:SPhiTh}
S_{\Phi,Thermal}=4kT\frac{Re(Z_{c})}{|Z_{c}|^{2}} \frac{M_{s}^{2}M_{p}^{2}}{(L_{i}+L_{p})^{2}}
\end{equation}
where $Z_{c}$ is the coupled resonator circuit impedance from (\ref{eq:Zcac}) and $T$ is the resonator temperature.

As long as all of these noise sources in the ac SQUID, including the intrinsic SQUID flux noise and amplifier noise in (\ref{eq:SPhiHEMT}), result in flux noise below the 100 mK thermal noise of the resonant circuit, as referred to a flux in the SQUID in (\ref{eq:SPhiTh}), the thermal noise can be resolved. As shown in the top right panel of Figure \ref{fig:acapp}, upon adjusting couplings to ensure hidden photon cavity Q of one million, we find that this is possible below $\approx 1$ GHz, when quantum noise in the resonator and SQUID begin to dominate.

\bibliography{white-paper_integrated}
\bibliographystyle{utphys}

\end{document}